\documentclass[aps,prb,showpacs,twocolumn,groupedaddress,subeqn,citesort]{revtex4}
\usepackage{latexsym}
\usepackage{graphicx}
\usepackage{amsmath}
\usepackage{amssymb}
\usepackage{bm}
\usepackage{float}
\usepackage{color}

\def\gtrless{\raise2.5pt\hbox{$>$}\llap{\lower2.5pt\hbox{$<$}}}
\def\gtrapprox{\raise2.5pt\hbox{$>$}\llap{\lower2.5pt\hbox{$\approx$}}}
\newcommand{\gl}[1]{Eq. (\ref{#1})}
\newcommand{\gls}[2]{Eqs. (\ref{#1},\ref{#2})}
\newcommand{\glto}[2]{Eqs. (\ref{#1}) to (\ref{#2})}
\newcommand{\bsq}[1]{\begin{subequations}\label{#1}}
\newcommand{\esq}{\end{subequations}}
\newcommand{\beq}[1]{\begin{equation}\label{#1}}
\newcommand{\eeq}{\end{equation}}
\newcommand{\beqa}[1]{\begin{eqnarray}\label{#1}}
\newcommand{\eeqa}{\end{eqnarray}}

\newcommand{\gd}{\dot{\gamma}}
\newcommand{\smop}{\Omega}
\newcommand{\smopb}{\Omega^{\dagger}}

\newcommand{\rb}{{\bf r}}

\newcommand{\qb}{{\bf q}}
\newcommand{\kb}{{\bf k}}
\newcommand{\pb}{{\bf p}}

\renewcommand{\rho}{\varrho}
\renewcommand{\epsilon}{\varepsilon}

\begin{document}

\title{Shear stresses of colloidal dispersions at the glass transition in equilibrium and in flow }
\author{J. J. Crassous, M. Siebenb{\"u}rger, and M. Ballauff}
\email[email:~]{Matthias.Ballauff@uni-bayreuth.de}
\affiliation{Physikalische Chemie I, University of Bayreuth, 95440
Bayreuth, Germany}
\author{M. Drechsler}
\affiliation{Makromolekulare Chemie II, University of Bayreuth,
95440 Bayreuth, Germany}
\author{D. Hajnal, O. Henrich, and  M. Fuchs}
\email[Email:~]{matthias.fuchs@uni-konstanz.de}
\affiliation{Fachbereich Physik, Universit\"at Konstanz,
 78457 Konstanz, Germany}

\date{\today}

\begin{abstract}
We consider a model dense colloidal dispersion at the glass
transition, and investigate the connection between equilibrium
stress fluctuations, seen in linear shear moduli, and the shear
stresses under strong flow conditions far from equilibrium,
viz.~flow curves for finite shear rates. To this purpose
thermosensitive core-shell particles consisting of a polystyrene
core and a crosslinked poly(N-isopropylacrylamide)(PNIPAM) shell
were synthesized. Data over an extended range in shear rates and
frequencies are compared to theoretical results from integrations
through transients and mode coupling approaches.  The connection
between non-linear rheology and glass transition is clarified. While
the theoretical models semi-quantitatively fit the data taken in
fluid states and the predominant elastic response of glass, a yet
unaccounted dissipative mechanism is identified in glassy states.
\end{abstract}
\pacs{82.70.Dd, 83.60.Df, 83.50.Ax, 64.70.Pf, 83.10.-y}

\maketitle

\section{Introduction}

Complex fluids and soft materials in general are characterized by a
strong variability in their rheological and elastic properties under
flow and deformations.\cite{Larson} Within the linear response
framework, storage- and loss- (shear) moduli describe elastic
contributions in solids and dissipative processes in fluids. Both
moduli are connected via Kramers-Kronig relations and result from
Fourier-transformations of a single time-dependent function, the
shear modulus $g^{\rm lr}(t)$. Importantly, the linear response
modulus $g^{\rm lr}(t)$ itself is defined in the  quiescent system
and (only) describes the small shear-stress fluctuations always
present in thermal equilibrium.\cite{Larson,russel}

Viscoelastic materials exhibit both, elastic and dissipative,
phenomena depending on external control parameters like temperature
and/ or density. The origins of the change between fluid and solid
like behavior can be manifold, including phase transitions of
various kinds. One mechanism existent quite universally in dense
systems is the glass transition, that structural rearrangements of
particles become progressively slower.\cite{Goe:92} It is
accompanied by a structural relaxation time which grows
dramatically. Maxwell was the first to describe this fluid-solid
transition phenomenologically. Dispersions consisting of colloidal,
slightly polydisperse (near) hard spheres arguably constitute one of
the most simple viscoelastic systems, where a glass transition has
been identified. It has been studied in detail by dynamic light
scattering
measurements,\cite{Pus:87,Meg:91,Meg:93,Meg:94,Heb:97,Bec:99,Bar:02,Eck:03}
confocal microscopy,\cite{Wee:00} linear,\cite{Mas:95,Zac:06} and
non-linear
rheology.\cite{Sen:99,Sen:99b,Pet:99,Pet:02,Pet:02b,Pha:06,Bes:07,Cra:06}
Computer simulations are available also.\cite{Phu:96,Str:99,Dol:00}
Mode coupling theory (MCT) has provided a semi-quantitative
explanation of the observed glass transition phenomena, albeit
neglecting ageing effects \cite{Pur:06} and decay processes at
ultra-long times that may cause (any) colloidal glass to flow
ultimately.\cite{Goe:91,Goe:92,Goe:99} Importantly, MCT predicts a
purely kinetic glass transition and describes it using only
equilibrium structural input, namely the equilibrium structure
factor $S_q$\cite{russel,dhont} measuring thermal density
fluctuations.

The stationary, nonlinear rheological behavior  under steady
shearing provides additional insight into the physics of dense
colloidal dispersions.\cite{Larson,russel} A priori it is not clear,
whether the mechanisms relevant during glass formation also dominate
the nonlinear rheology. Solvent mediated interactions (hydrodynamic
interactions), which do not affect the equilibrium phase diagram,
may become crucially important. Also, shear may cause ordering or
layering of the particles.\cite{Lau:92} Simple phenomenological
relations between the frequency dependence of the linear response
and the shear rate dependence of the nonlinear response, like the
Cox-Merz rule, have been formulated, but often lack firm theoretical
support or are limited to special shear
histories.\cite{Larson,Wys:07}

On the other hand, within a number of theoretical approaches a
connection between steady state rheology and the glass transition
has been suggested. Brady worked out a scaling description of the
rheology based on the concept that the structural relaxation arrests
at random close packing.\cite{Bra:93} In the  soft glassy rheology
model, the trap model of glassy relaxation by Bouchaud was
generalized to describe mechanical deformations and
ageing.\cite{Sol:97,Sol:98,Fie:00} The mean field approach to  spin
glasses was generalized to systems with broken detailed balance in
order to model flow curves of glasses under
shear.\cite{Ber:00,Ber:02} The application of these novel approaches
to colloidal dispersions has lead to numerous insights, but has been
hindered by the use of unknown parameters in the approaches.  MCT,
also, was generalized to include effects of
shear,\cite{Miy:02,miyazaki,Kob:05} and, within the {\it
integrations through transients} (ITT) approach, to quantitatively
describe all aspects of stationary states under steady
shearing.\cite{Fuc:02,Fuc:03,Fuc:05c} Some aspects of the ITT
approach to flow curves have already been
tested,\cite{Cra:06,Var:06} but the connection, central in the
approach, between fluctuations around equilibrium and the nonlinear
response, has not been investigated experimentally up to now.

In the present contribution we explore the connection between
structural relaxation close to glassy arrest and the rheological
properties far from equilibrium. Thereby we crucially test the ITT
approach, which aims to unify the understanding of these phenomena.
It requires, as sole input, information on the equilibrium structure
(namely $S_q$), and, first gives a formally exact generalization of
the shear modulus to finite shear rates, $g(t,\gd)$, which is then
approximated in a consistent way. We investigate  a model dense
colloidal dispersion at the glass transition, and determine its
linear and nonlinear rheology. Thermosensitive core-shell particles
consisting of a polystyrene core and a crosslinked
poly(N-isopropylacrylamide)(PNIPAM) shell were synthesized and their
dispersions characterized in detail.\cite{Cra:06,Cra:07} Data over
an extended range in shear rates and frequencies are compared to
theoretical results from MCT and ITT.

The paper is organized as follows: section II summarizes the
equations of the microscopic ITT approach in order to provide a
self-contained presentation of the theoretical framework. In section
III some of the universal predictions of ITT are discussed in order
to describe the phenomenological properties of the non-equilibrium
transition studied in this work. Building on the universal
properties, section IV introduces a simplified model  which
reproduces the phenomenology. Section V introduces the experimental
system. Section VI contains the main part of the present work, the
comparison of combined measurements of the linear and non-linear
rheology of the model dispersion with calculations in microscopic
and simplified theoretical models. A short summary concludes the
paper, while the appendix presents an extension of the simplified
model used in the main text.

\section{Microscopic approach}
\label{model}

We consider $N$ spherical particles with radius $R_H$ dispersed in a
volume $V$ of solvent (viscosity $\eta_s$) with imposed homogeneous,
and constant linear shear-flow. The flow velocity points along the
$x$-axis and its gradient along the $y$-axis. The motion of  the
particles (with positions $\rb_i(t)$ for $i=1,\ldots,N$) is
described by  $N$ coupled Langevin equations\cite{dhont} \beq{b0}
\zeta \left( \frac{d \rb_i  }{d t}  - {\bf v}^{\rm solv}(\rb_i)
\right)
 = {\bf F}_i + {\bf f}_i\; .
\eeq Solvent friction is measured by the Stokes friction coefficient
$\zeta=6\pi\eta_s R_H$. The $N$  vectors ${\bf
F}_i=-\partial/\partial \rb_i\, U(\left\{\rb_j\right\})$ denote the
interparticle force on particle $i$ deriving from potential
interactions with all other particles; $U$ is the  potential energy
which depends on all particles' positions. The solvent shear-flow is
given by ${\bf v}^{\rm solv}(\rb)=\gd\, y\, \hat{\bf x}$, and the
Gaussian white noise force satisfies (with $\alpha,\beta$ denoting
directions)
$$\langle f^\alpha_i(t)\; f^\beta_j(t') \rangle = 2 \zeta \, k_BT\,
\delta_{\alpha \beta}\, \delta_{ij}\, \delta(t-t')\; ,$$ where
$k_BT$ is the thermal energy. Each particle experiences
interparticle forces, solvent friction, and random  kicks.
Interaction  and friction forces on each particle balance on
average, so that the particles are at rest in the solvent on
average. The Stokesian friction is proportional to the particle's
motion {\em relative to} the solvent flow at its position; the
latter varies linearly with $y$. The random force on the level of
each particle satisfies the fluctuation dissipation relation.

An important approximation in \gl{b0} is the neglect of hydrodynamic
interactions, which would arise from the proper treatment of the
solvent flow around moving particles.\cite{russel,dhont} In the
following we will argue that such effects can be neglected at high
densities where interparticle forces hinder and/or prevent
structural rearrangements, and where the system is close to arrest
into an amorphous, metastable solid. Another important approximation
in \gl{b0} is the assumption of a given, constant shear rate $\gd$,
which does not vary throughout the (infinite) system. We start with
this assumption in the philosophy that, first, homogeneous states
should be considered, before heterogeneities and confinement effects
are taken into account. All difficulties in \gl{b0} thus are
connected to the many-body interactions given by the forces ${\bf
F}_i$, which couple the $N$ Langevin equations. In the absence of
interactions, ${\bf F}_i\equiv0$, \gl{b0} leads to  super-diffusive
particle motion termed 'Taylor dispersion'.\cite{dhont}

While formulation of the considered microscopic model handily uses
Langevin equations, theoretical analysis proceeds more easily from
the reformulation of \gl{b0} as Smoluchowski equation. It describes
the temporal evolution of the distribution function
$\Psi(\left\{\rb_i\right\} ,t)$ of the particle positions
\beq{b01}
\partial_t \Psi(\left\{\rb_i\right\} ,t)  = \smop \; \Psi(\left\{\rb_i\right\} ,t)\; ,
\eeq
employing the Smoluchowski operator,\cite{russel,dhont}
\beq{b1} \smop =  \sum_{j=1}^{N} \left[ D_0\;
\frac{\partial}{\partial \rb_j} \cdot \left(
\frac{\partial}{\partial \rb_j}  - \frac{1}{k_BT}\, {\bf F}_j
\right) - \gd\, \frac{\partial}{\partial x_j}\, y_j \right] \;, \eeq
built with the (bare) diffusion coefficient $D_0=k_BT/\zeta$ of a
single particle. We assume that the system relaxes into a unique
stationary state at long times, so that $\Psi(t\to\infty)=\Psi_s$
holds. Homogeneous, amorphous systems are studied so that the
stationary distribution function $\Psi_s$ is translationally
invariant but anisotropic. Neglecting ageing, the  formal solution
of the Smoluchowski equation within the ITT approach can be brought
into the form\cite{Fuc:02,Fuc:05c}
\beq{b2} \Psi_s =\Psi_e + \frac{\gd}{k_BT} \; \int_0^\infty\!\!\!\!
dt \;\Psi_e \; \sigma_{xy} \; e^{\smopb t }\; , \eeq
where the adjoint Smoluchowski $\smopb$ operator arises from partial
integrations. It acts on the quantities to be averaged with
$\Psi_s$. $\Psi_e$ denotes the equilibrum canonical distribution
function, $\Psi_e\propto e^{-U/(k_BT)}$, which is the
time-independent solution of \gl{b01} for $\gd=0$; in \gl{b2}, it
gives the initial distribution at the start of shearing (at $t=0$).
The potential part of the stress tensor
\mbox{$\sigma_{xy}=-\sum_{i=1}^N F^x_i\,y_i$} entered via $\smop
\Psi_e = \gd\, \sigma_{xy}\, \Psi_e$. The simple, exact result
\gl{b2} is central to the ITT approach as it connects steady state
properties to time integrals formed with the shear-dependent
dynamics. The latter contains slow intrinsic particle motion.

\begin{figure}[t]
\centering
\includegraphics[ width=0.6\columnwidth]{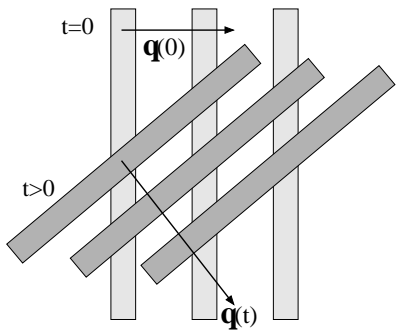}
\caption{Shear advection of a fluctuation with initial wavevector in
$x$-direction, $\qb(t\!\!=\!\!0)=q\, (1,0,0)^T$, and advected
wavevector at later time  $\qb(t\!\!>\!\!0)= q\,  (1,-\gd t,0)^T$.
At all times, $\qb(t)$ is perpendicular to the planes of constant
fluctuation amplitude. Note that the magnitude $q(t)=q\sqrt{1+(\gd
t)^2}$ increases with time. Brownian motion, neglected in this
sketch, would smear out the fluctuation.\label{Fig0}}
\end{figure}

In ITT, the evolution towards the stationary distribution at
infinite times is approximated by following the slow structural
rearrangements, encoded in the transient density correlator
$\Phi_\qb(t)$. It is defined by\cite{Fuc:02,Fuc:05c}
\beq{b3} \Phi_\qb(t) = \frac{1}{NS_q}\; \langle\,
\delta\varrho^*_{\qb} \; e^{\smopb t } \; \delta\varrho_{\qb(t)}\,
\rangle^{(\gd=0)}\; . \eeq
It describes the fate of an equilibrium density fluctuation with
wavevector $\qb$, where $\varrho_\qb = \sum_{j=1}^N e^{i \qb \cdot
\rb_j}$, under the combined effect of internal forces, Brownian
motion and shearing. Note that because of the appearance of $\Psi_e$
in \gl{b2}, the average in \gl{b3} can be evaluated with the
equilibrium canonical distribution function, while the dynamical
evolution contains Brownian motion and shear advection. The
normalization is given by $S_q$ the equilibrium structure factor
\cite{russel,dhont} for wavevector modulus $q=|{\bf q}|$. The {\em
advected} wavevector enters in \gl{b3} \beq{b4} {\bf q}(t) =  {\bf
q} - \gd t\; q_x \; \hat{\bf y}\; ,
 \eeq
where unit-vector $\hat{\bf y}$ points in $y$-direction) The
time-dependence in ${\bf q}(t)$ results from the affine particle
motion with the shear flow of the solvent. Translational invariance
under shear dictates that at a time $t$ later, the equilibrium
density fluctuation $\delta\varrho^*_\qb$ has a nonvanishing overlap
only with the advected fluctuation $\delta\varrho_{\qb(t)}$; see
Fig. \ref{Fig0}, where a non-decorrelating fluctuation is sketched
under shear. In the case of vanishing Brownian motion, viz.~$D_0=0$
in \gl{b1}, we find $\Phi_{\qb}(t)\equiv1$, because the advected
wavevector takes account of simple affine particle
motion.\cite{achtung} The relaxation of $\Phi_{\qb}(t)$ thus heralds
decay of structural correlations. Within ITT, the time integral over
such structural decorrelations provides an approximation to the
stationary state:
\beqa{b5} \Psi_s &\approx& \Psi_e +  \\ &&
\frac{\gd}{2k_BT}\int_0^\infty\!\!\!\!\!\!dt \sum_\kb \frac{k_xk_y\,
S'_{k}}{k\, N S^2_{\kb(t)}} \; \Phi^2_{\kb}(t) \; \left( \Psi_e \,
\varrho^*_{\kb(t)} \, \varrho_{\kb(t)} \right)\,,\nonumber \eeqa
with  $S'_k=\partial S_k/\partial k$.\cite{eq5com}  The last term in
brackets in \gl{b5} expresses, that the expectation value of a
general fluctuation $ A$ in ITT-approximation contains the
(equilibrium) overlap with the local structure,  $\langle
\varrho^*_\kb \, \varrho_\kb\, A \rangle^{(\gd=0)}$. The difference
between the equilibrium and stationary distribution functions then
follows from integrating over time the spatially resolved
(viz.~wavevector dependent) density variations.

The general results for $\Psi_s$, the exact one of \gl{b2} and the
approximation \gl{b5}, can be applied to compute  stationary
expectation values like for example the thermodynamic transverse
stress, $\sigma(\gd)=\langle\sigma_{xy}\rangle /V$. Equation
(\ref{b2}) leads to  an exact non-linear Green-Kubo relation:
\beq{b55}\sigma(\gd)= \gd \int_0^\infty dt\; g(t,\gd)\; , \eeq where
the generalized shear modulus $g(t,\gd)$ depends on shear rate via
the Smoluchowski operator from \gl{b1} \beq{b6} g(t,\gd)=
\frac{1}{k_BT V}\; \langle\, \sigma_{xy} \; e^{\smopb t } \;
\sigma_{xy}\, \rangle^{(\gd=0)} \; . \eeq In ITT, the slow stress
fluctuations in $g(t,\gd)$ are approximated by following the slow
structural rearrangements, encoded in the transient density
correlators. The generalized modulus becomes, using the
approximation \gl{b5}, or, equivalently, performing a mode coupling
approximation:\cite{Fuc:03,Fuc:02,miyazaki}
\beq{b7} g(t,\gd) = \frac{k_BT}{2} \,
\int\!\!\frac{d^3k}{(2\pi)^3}\; \frac{k_x^2k_yk_y(t)}{k\, k(t)}\;
\frac{S'_kS'_{k(t)}}{S^2_{k(t)}}\; \Phi^2_{\kb}(t)\; , \eeq
Summation over wavevectors has been turned into integration in
\gl{b7} considering an infinite system.

The familiar shear modulus of linear response theory describes
thermodynamic stress fluctuations in equilibrium, and is obtained
from \gls{b6}{b7} by setting $\gd=0$.\cite{Larson,russel,Nae:98}
While \gl{b6} then gives the exact Green-Kubo relation, the
approximation \gl{b7} turns into the well-studied MCT formula. For
finite shear rates, \gl{b7} describes how affine particle motion
causes stress fluctuations to explore shorter and shorter length
scales. There the effective forces, as measured by the gradient of
the direct correlation function, $S'_k/S_k^2 = n c'_k=n\partial
c_k/\partial k$, become smaller, and vanish asympotically,
$c'_{k\to\infty} \to 0$; the direct correlation function $c_{k}$ is
connected to the structure factor via the Ornstein-Zernicke equation
$S_k=1/(1-n\, c_k)$, where $n=N/V$ is the particle density. Note,
that the equilibrium structure suffices to quantify the effective
interactions, while shear just pushes the fluctuations around on the
'equilibrium energy landscape'.

Structural rearrangements of the dispersion affected by Brownian
motion is encoded in the transient density correlator. Shear induced
affine motion, viz.~ the case $D_0=0$, is not sufficient to cause
$\Phi_\kb(t)$ to decay. Brownian motion of the quiescent correlator
$\Phi^{(\gd=0)}_{k}(t)$ leads at high densities to a slow structural
process which arrests at long times in (metastable) glass states.
Thus the combination of structural relaxation and shear is
interesting. The interplay between intrinsic structural motion and
shearing in $\Phi_\kb(t)$ is  captured by $(i)$ first a  formally
exact Zwanzig-Mori type equation of motion, and $(ii)$ second a mode
coupling factorisation in the memory function built with
longitudinal stress fluctuations.\cite{Fuc:02,Fuc:05c}  The equation
of motion for the transient density correlators is
\beq{b8}
\partial_t \Phi_\qb(t) + \Gamma_\qb(t) \; \left\{
\Phi_\qb(t) + \int_0^t dt'\; m_\qb(t,t') \; \partial_{t'}\,
\Phi_\qb(t') \right\} = 0 \; , \eeq
where the initial decay rate
$\Gamma_\qb(t)= D_0\, q^2(t)/S_{q(t)}$ generalizes the familiar
result from linear response theory to advected wavevectors; it contains Taylor dispersion. The
memory equation contains fluctuating stresses and similarly like
$g(t,\gd)$ in \gl{b5}, is calculated in mode coupling approximation
 \beq{b9} m_\qb(t,t') = \frac{1}{2N} \sum_{\kb}
V_{\qb\kb\pb}(t,t')\;
 \Phi_{\kb(t')}(t-t')\;  \Phi_{\pb(t')}(t-t')\; ,
\eeq where we abbreviated $\pb=\qb-\kb$. The vertex generalizes the
expression in the quiescent case:\cite{Fuc:02}
\beqa{b10} V_{\qb\kb\pb}(t,t') &=& \frac{S_{\qb(t)} \, S_{\kb(t')}\,
S_{\pb(t')}}{q^2(t)\, q^2(t')}\;
{\cal V}_{\qb\kb\pb}(t)\; {\cal V}_{\qb\kb\pb}(t')\, , \nonumber \\
{\cal V}_{\qb\kb\pb}(t) &=& \qb(t)\cdot \left( \, \kb(t)\; n
c_{k(t)} + \pb(t)\; n c_{p(t)}\, \right)\;.  \eeqa
With shear, wavevectors in \gl{b10} are advected according to
\gl{b4}.

Equations (\ref{b5},\ref{b8},\ref{b9}), with the specific example of
the generalized shear modulus \gl{b7}, form a closed set of
equations determining  rheological properties of a sheared
dispersion from equilibrium structural input.\cite{Fuc:02,Fuc:05c}
Only the static structure factor $S_q$ is required to predict $(i)$
the time dependent shear modulus within linear response, $g^{\rm
lr}(t)= g(t,\gd=0)$, and $(ii)$ the stationary stress $\sigma(\gd)$
from \gl{b55}. The loss and storage moduli of small amplitude
oscillatory shear measurements\cite{Larson,russel} follow from
\gl{b6} in the linear response case $(i)$
\beq{b11} G'(\omega) + i\, G''(\omega) = i \omega\; \int_0^\infty
dt\; e^{-i\, \omega\, t}\; g(t,\gd=0) \; . \eeq
While, in the linear response regime, modulus and density correlator
are measurable quantities, outside the linear regime, both
quantities serve as tools in the ITT approach only. The transient
correlator and shear modulus provide a route to the stationary
averages, because they describe the decay of equilibrium
fluctuations under external shear, and their time integral provides
an approximation for the stationary distribution function, see
\gl{b5}. Determination of the frequency dependent moduli under large
amplitude oscillatory shear has become possible recently
only,\cite{Miy:06} and requires an extension of the present approach
to time dependent shear rates in \gl{b1}.\cite{Bra:07}

\section{Universal aspects}

\begin{figure}[t]
\centering
\includegraphics[ width=0.8\columnwidth]{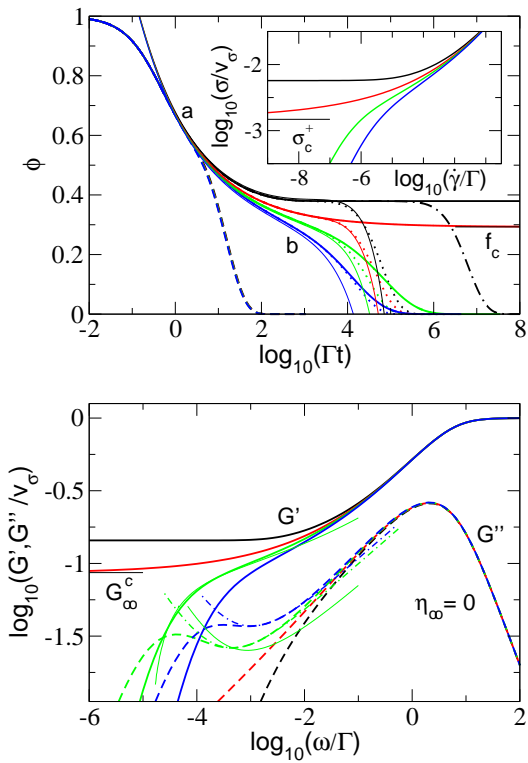}

\caption{Overview of the properties of the F$_{12}^{(\gd)}$-model
characteristic for the transition between  fluid and yielding glass.
The upper panel shows numerically obtained transient correlators
$\Phi(t)$ for $\epsilon=0.01$ (black curves),   $\epsilon=0$ (red),
$\epsilon=-0.005$  (green), and $\epsilon=-0.01$  (blue). The shear
rates are $\left|\dot{\gamma}/\Gamma\right|=0$ (thick solid lines),
$\left|\dot{\gamma}/\Gamma\right|=10^{-6}$ (dotted lines), and
$\left|\dot{\gamma}/\Gamma\right|=10^{-2}$ (dashed lines). For the
glass state at $\epsilon=0.01$ (black),
$\left|\dot{\gamma}/\Gamma\right|=10^{-8}$ (dashed-dotted-line) is
also included. All curves were calculated with  $\gamma_{c}=0.1$ and
$\eta_{\infty}=0$. The thin solid lines give the factorization
result \gl{c1} with scaling functions $\cal G$  for
$\left|\dot{\gamma}/\Gamma\right|=10^{-6}$;  label $a$ marks the
critical law (\ref{c3}), and label $b$ marks the von Schweidler-law
(\ref{c4}).  The critical glass form factor $f_c$ is indicated. The
inset shows the flow curves for the same values for $\epsilon$. The
thin black bar shows the yield stress $\sigma^+_c$ for $\epsilon=0$.
The lower panel shows the viscoelastic storage (solid line) and loss
(broken line  modulus  for the same values of $\epsilon$. The thin
green lines are the Fourier-transformed factorization result \gl{c1}
with scaling function $\cal G$ taken from the upper panel for
$\epsilon=-0.005$. The dashed-dotted lines show the  fit formula
\gl{e1} for the spectrum in the minimum-region with $G_{\rm
min}/v_{\sigma}=0.0262$, $\omega_{\rm min}/\Gamma=0.000457$ at
$\epsilon=-0.005$ (green) and $G_{\rm min}/v_{\sigma}=0.0370$,
$\omega_{\rm min}/\Gamma=0.00105$ at $\epsilon=-0.01$ (blue). The
elastic constant at the transition $G_\infty^c$ is marked also,
while the high frequency asymptote $G'_\infty=G'(\omega\to\infty)$
is not labeled explicitly. \label{Figneu} }
\end{figure}

The summarized microscopic ITT equations contain a bifurcation in
the long-time behavior of $\Phi_\qb(t)$,  which corresponds to a
non-equilibrium transition between a fluid and a shear-molten glassy
state; it is described in this section. Close to the
transition, (rather) universal predictions can be made about the
non-linear dispersion rheology and the steady state properties. The
central predictions are introduced in this section and summarized in
the overview figure \ref{Figneu}. It is obtained from the schematic
model which is also used to analyse the data, and which is
introduced in the following Sect. IV.

A dimensionless separation parameter $\epsilon$ measures the
distance to the transition which is situated at $\epsilon=0$. A
fluid state ($\epsilon < 0$) possesses a (Newtonian) viscosity,
$\eta_0(\epsilon<0) =\lim_{\gd\to0} \sigma(\gd)/\gd$, and shows
shear-thinning upon increasing $\gd$. Via the relation
$\eta_0=\lim_{\omega\to0}\, G''(\omega)/\omega$, the Newtonian
viscosity can also be taken from the loss modulus at low
frequencies, where $G''(\omega)$ dominates over the storage modulus.
The latter varies like $G'(\omega\to0)\sim \omega^2$. A glass
($\epsilon\ge0$), in the absence of flow, possesses an elastic
constant $G_\infty$, which can be measured in the elastic shear
modulus $G'(\omega)$ in the limit of low frequencies,
$G'(\omega\to0,\epsilon\ge0)\to G_\infty(\epsilon)$. Here the
storage modulus dominates over the loss one, which drops like
$G''(\omega\to0)\sim \omega$. (Note that the high frequency
modulus $G'_\infty=G'(\omega\to\infty)$ is characteristic of the
particle interactions,\cite{Lio:94} and exists in fluid and solid
states.) Enforcing steady shear flow melts the glass. The
stationary stress of the shear-molten glass always exceeds a
(dynamic) yield stress. For decreasing shear rate, the viscosity
increases like $1/\gd$, and the stress levels off onto the
yield-stress plateau,
$\sigma(\gd\to0,\epsilon\ge0)\to\sigma^+(\epsilon)$.

Close to the transition, the zero-shear viscosity $\eta_0$, the
elastic constant $G_\infty$, and the yield stress $\sigma^+$ show
universal  anomalies as functions of the distance to the transition:
the viscosity diverges in a power-law $\eta_0(\epsilon\to0-) \sim
(-\epsilon)^{-\gamma}$ with material dependent exponent $\gamma$
around $2-3$, the elastic constant increases like a square-root
$G_\infty(\epsilon\to0+)-G_\infty^c \sim \sqrt{\epsilon}$, and the
dynamic yield stress $\sigma^+(\epsilon\to0+)$ also increases with
infinite slope above its value $\sigma^+_c$ at the bifurcation. The
quantities $G_\infty^c$ and $\sigma^+_c$ denote the respective
values at the transition point $\epsilon=0$, and measure the jump in
the elastic constant   and in the yield stress  at the glass
transition; in the fluid state,  $G_\infty(\epsilon<0)=0$ and
$\sigma^+(\epsilon<0)=0$ hold.

The described results follow from the stability analysis of
\gls{b8}{b9} around an arrested, glassy structure $f_q$ of the
transient correlator.\cite{Fuc:02,Fuc:03} Considering the time
window where $\Phi_\qb(t)$ is metastable and close to arrest at
$f_q$, and taking all control parameters like density, temperature,
etc. to be close to the values at the transition, the stability
analysis yields the 'factorization' between spatial and temporal
dependences \beq{c1} \Phi_\qb(t) = f^c_q + h_q \; {\cal
G}(\,t/t_0,\epsilon,\gd t_0\,) + \ldots\; , \eeq where the
(isotropic) glass form factor $f^c_q$ and critical amplitude $h_q$
describe the spatial properties of the metastable glassy state. The
critical glass form factor $f^c_q$ gives the long-lived component of
density fluctuations, and $h_q$ captures local particle
rearrangements. Both can be taken as constants independent on shear
rate and density, as they are evaluated from the vertices in
\gl{b10} at the transition point. All time-dependence and
(sensitive) dependence on the external control parameters is
contained in the function $\cal G$, which often is called
'$\beta$-correlator' and obeys the non-linear stability equation
\beq{c2} \epsilon - c^{(\gd)}\; (\gd t)^2 + \lambda\; {\cal G}^2(t)
= \frac{d}{dt} \int_0^t\!\!\!dt'\; {\cal G}(t-t')\; {\cal G}(t')\; ,
\eeq with initial condition \beq{c3} {\cal G}(t \to 0) \to
(t/t_0)^{-a} \; . \eeq The two parameters $\lambda$ and $c^{(\gd)}$
in \gl{c2} are determined by the static structure factor at the
transition point, and take values around $\lambda\approx 0.73$ and
$c^{(\gd)}\approx 3$ for $S_q$ taken from Percus-Yevick
approximation\cite{russel} for hard sphere
interactions.\cite{Fuc:02,Fuc:03,ishsm} The transition point then
lies at packing fraction $\phi_c=\frac{4\pi}{3} n_c R_H^3\approx
0.52$ (index $c$ for critical), and the separation parameter
measures the relative distance,  $\epsilon=C\, (\phi-\phi_c)/\phi_c$
with $C \approx 1.3$. The 'critical' exponent $a$ is given by the
exponent parameter $\lambda$ via
$\lambda=\Gamma(1-a)^2/\Gamma(1-2a)$.\cite{Goe:91,Goe:92}

The time scale $t_0$ in \gl{c3} provides the means to match the
function ${\cal G}(t)$ to the microscopic, short-time dynamics. The
\gls{b8}{b9} contain a simplified description of the short time
dynamics in colloidal dispersions via the initial decay rate
$\Gamma_\qb(t)$.  From this model for the short-time dynamics, the
time scale $t_0\approx 1.6\, 10^{-2} R_H^2/D_0$  is obtained.
Solvent mediated effects on the short time dynamics are well known
and are neglected in $\Gamma_\qb(t)$ in \gl{b8}. Within the ITT
approach, one finds that if hydrodynamic interactions were included
in \gl{b8}, all of the mentioned universal predictions 
 would remain true. Only the value of $t_0$ will be
shifted and depend on the short time hydrodynamic interactions.
For the quiescent glass transition this has been discussed
within MCT,\cite{Fra:98} and ITT extends this to driven cases.
This statement remains valid in ITT, as long as the
hydrodynamic interactions do not affect the mode coupling vertex in
\gl{b10}. In this sense, hydrodynamic interactions can be
incorporated into the theory of the glass transition, and amount to
a rescaling of the matching time $t_0$, only.

Obviously, the matching time $t_0$ also provides an upper cut-off
for the time window of the structural relaxation. At times shorter
than $t_0$ the specific short-time dynamics matters. The condition
$\gd t_0\ll 1$ follows and translates into a restriction for the
accessible range of shear rates, $\gd \ll \gd_*$, where the
upper-cut off shear rate $\gd_*$ is connected to the matching time.

The parameters $\epsilon$, $\lambda$ and $c^{(\gd)}$ in \gl{c2} can
be determined from the equilibrium structure factor $S_q$ at or
close to the transition, and, together with $t_0$ and the shear rate
$\gd$ they capture the essence of the rheological anomalies in dense
dispersions. A divergent viscosity follows from the prediction of a
strongly increasing final relaxation time in $\cal G$ in the
quiescent fluid phase \beq{c4} {\cal G}(t \to \infty , \epsilon < 0,
\gd = 0 ) \to - \left(t / \tau \right)^{b} \quad ,\;\; \mbox{with
}\; \frac{t_0}{\tau} \propto (-\epsilon)^\gamma\; . \eeq The
entailed temporal power law, termed von Schweidler law, initiates
the final decay of the correlators, which has a density and
temperature independent shape $\tilde \Phi_q(\tilde t)$.  The final
decay, often termed $\alpha$-relaxation, depends on $\epsilon$ only
via the time scale $\tau(\epsilon)$ which rescales the time, $\tilde
t=t/\tau$. Equation (\ref{c2}) establishes the crucial time scale
separation between $t_0$ and $\tau$, the divergence of $\tau$, and
the stretching (non-exponentiality) of the final decay; it also
gives the values of the exponents via
$\lambda=\Gamma(1+b)^2/\Gamma(1+2b)$, and $\gamma=(1/a+1/b)/2$.
Using \gl{b7}, the divergence of the Newtonian viscosity
follows.\cite{Goe:92,Goe:91} During the final decay the shear
modulus becomes a function of rescaled time, $\tilde g(\tilde
t=t/\tau,\gd=0)$, leading to $\eta_0 \propto \tau(\epsilon)$; its
initial value is given by the elastic constant at the transition,
$\tilde g(\tilde t \ll 1,\epsilon\to0-,\gd=0) = G_\infty^c$.

On the glassy side of the transition, $\epsilon\ge 0$, the transient
density fluctuations stays close to a plateau value for intermediate
times which increases when going deeper into the glass,
\beq{c5}
{\cal G}(t_0 \ll t \ll 1/|\gd| , \epsilon \ge 0 ) \to
\sqrt{\frac{\epsilon}{1-\lambda}} + {\cal O}(\epsilon)\; . \eeq
Entered into \gl{b7}, the square-root dependence of the plateau
value translates into the square-root anomaly of the elastic
constant $G_\infty$, and causes the increase of the yield stress close to the glass transition.

Only, for vanishing shear rate, $\gd=0$, an ideal glass state exists
in the ITT approach for steady shearing. All density correlators
arrest at a long time limit, which from \gl{c5} close to the
transition is given by
$\Phi_\qb(t\to\infty,\epsilon\ge0,\gd=0)=f_q=f^c_q + h_q
\sqrt{\epsilon/(1-\lambda)} +{\cal O}(\epsilon)$. Consequently the
modulus remains elastic at long times,
$g(t\to\infty,\epsilon\ge0,\gd=0)=G_\infty>0$. Any (infinitesimal)
shear rate, however, melts the glass and causes a final decay of the
transient correlators. The function $\cal G$ initiates the decay
around the critical plateau of the transient correlators and sets
the common time scale for the final decay under shear
\beq{c6} {\cal G}(t \to \infty , \epsilon \ge 0 ) \to -
\sqrt{\frac{c^{(\gd)}}{\lambda-\frac 12}} \; |\gd t| \; . \eeq
Under shear  all correlators decay from the plateau as function of
$|\gd t|$. Steady shearing thus prevents non-ergodic arrest and
restores ergodicity. This aspect of \gl{c2} has two important
ramifications for the steady state of shear molten
glasses.\cite{Fuc:02,Fuc:03} First, ITT finds that shear melts a
glass and produces a unique steady state at long times. This
conclusion is restricted by the assumption of homogeneous states and
excludes the possible existence of ordering or layering under shear.
Also, ageing was neglected, which could remain because of
non-ergodicity in the initial quiescent state. (Ergodicity of the
sheared state however suggests ageing to be unimportant under
shear.\cite{Ber:00,Fie:00}) Second, all stationary averages, which
in ITT are obtained from integrating up the transient fluctuations,
do not exhibit a linear response regime in the glass. Rather they
take finite values for vanishing shear rate, which jump
discontinuously at the glass transition. This holds because the
shear-driven decay of \gl{c6} initiates a scaling law where the
transient correlators decay as function of $|\gd t|$ down from the
plateau $f_q$ to zero, denoted as $\Phi^+_\qb(t|\gd |)$. When
entered into \gl{b5}, time appears only in the combination together
with shear rate and thus after time integration the shear rate
dependence drops out, yielding a finite result even in the limit of
infinitesimal shear rate. Prominent example of a stationary average
that has no linear response regime with respect to $\gd$  in the
glass phase is the shear stress $\sigma(\gd,\epsilon\ge0)$. It takes
finite values for vanishing shear rate, $\sigma^+(\epsilon)=
\sigma(\gd\to0,\epsilon\ge0)$, and jumps at the glass transition
from zero to a finite value. Because of \gl{c5} it increases rapidly
when moving deeper into the glass.

\section{Schematic model}

The universal aspects described in the previous section are
contained in any ITT model that contains the central bifurcation
scenario and recovers \gls{c1}{c2}. Equation (\ref{c1}) states that
spatial and temporal dependences decouple in the intermediate time
window. Thus it is possible to investigate ITT models without proper
spatial resolution. Because of the technical difficulty to evaluate
the anisotropic functionals in \gls{b7}{b9}, it is useful to
restrict the description to few or to a single transient correlator.
In the schematic $F_{12}^{(\dot{\gamma})}$-model,\cite{Fuc:03} a
single 'typical' density correlator $\Phi(t)$, conveniently
normalized according to $\Phi(t\to0)= 1-\Gamma t$, obeys a
Zwanzig-Mori memory equation which is modeled according to \gl{b8}
\begin{equation}\label{d1}
\partial_t \Phi(t) + \Gamma \left\{ \Phi(t) + \int_0^tdt'\; m(t-t') \;
\partial_{t'} \Phi(t') \right\}  = 0 \; .
\end{equation}
The parameter $\Gamma$ mimics the microscopic dynamics of the
'typical' density correlator chosen in \gl{d1}, and will depend on
structural and hydrodynamic correlations. The memory function
describes stress fluctuations which become more sluggish together
with density fluctuations, because slow structural rearrangements
dominate all quantities. A self consistent approximation closing the
equations of motion  is made mimicking \gl{b9}. In the
$F_{12}^{(\dot{\gamma})}$-model one includes a linear term (absent
in \gl{b9}) in order to $(i)$ sweep out the full range of $\lambda$
values in \gl{c2}, and $(ii)$ retain algebraic simplicity: \beq{d2}
m(t)= \frac{v_1 \, \Phi(t) + v_2\, \Phi^2(t)}{1+\left(\dot\gamma
t/\gamma_c\right)^2} \eeq

This model, for the quiescent case $\dot\gamma=0$, had been
suggested by G\"otze in 1984,\cite{Goe:84,Goe:91} and describes the
development of slow structural relaxation upon increasing the
coupling vertices $v_i\ge0$; they mimic the dependence of the
vertices  in \gl{b9} at $\gd=0$ on the equilibrium structure given
by $S_q$. Under shear an explicit time dependence of the couplings
in $m(t)$ captures the accelerated loss of memory by shear advection
in \gl{b9}. Shearing causes the dynamics to decay for long times,
because fluctuations are advected to smaller wavelengths where small
scale Brownian motion relaxes them. Equations (\ref{d1},\ref{d2})
lead, with $\Phi(t)=f^c+(1-f^c)^2\,{\cal G}(t,\epsilon,\gd)$, and
the choice of the vertices $v_2=v_2^c=2$, and
$v_1=v_1^c+\epsilon\,(1-f^c)/f^c$, where $v_1^c=0.828$,   to the
critical glass form factor $ f^c= 0.293$ and to the stability
equation (\ref{c2}), with parameters
$$\lambda= 0.707\, ,\;  c^{(\gd)}= 0.586/\gamma_c^2\, \mbox{, and }\; t_0= 0.426 /\Gamma \;.
$$
 The choice of transition point
$(v_1^c,v_2^c)$ is motivated by its repeated use in the literature.
Actually, there is a line of glass transitions where the long time
limit $f=\Phi(t\to\infty)$ jumps discontinuously. It is
parameterized by $(v_1^c,v_2^c)=((2\lambda-1),1)/\lambda^2$ with
$0.5\le \lambda< 1$, and $f^c=1-\lambda$. The present choice is just a typical one,
which corresponds to the given typical $\lambda$-value. The
separation parameter $\epsilon$ is the crucial control parameter as
it takes the system through the transition. The parameter $\gamma_c$
is a scale for the magnitude of strain that is required in order for
the accumulated strain $\gd t$ to matter.\cite{erweiterung} In
\gl{c2}, it is connected to the parameter $c^{(\gd)}$ .

For simplicity, the quadratic dependence of the generalized shear
modulus on density fluctuations is retained from the microscopic
\gl{b7}. It simplifies because only one density mode is considered,
and as, for simplicity, a possible dependence of the vertex
(prefactor) $v_\sigma$ on shear is neglected
\beq{d4}
g(t)= v_\sigma \,\Phi^2(t) + \eta_\infty \; \delta(t) \; .
 \eeq
The parameter $\eta_\infty$ characterizes a short-time, high
frequency viscosity and models viscous processes which require no
structural relaxation. Together with $\Gamma$ (respectively $t_0$),
it is the only model parameter affected by solvent mediated
interactions. Steady state shear stress under constant shearing, and
viscosity then follow via integrating up the generalized modulus:
\beq{d5}
 \sigma = \eta \; \dot\gamma =  \dot{\gamma}\; \int_0^\infty\!\!\!
dt\; g(t)  = \dot{\gamma}\; \int_0^\infty\!\!\!dt\;v_\sigma
\Phi^2(t) + \dot{\gamma}\; \eta_\infty \; . \eeq Also, when setting
shear rate $\gd=0$ in \gls{d1}{d2}, so that the schematic correlator
belongs to the quiescent, equilibrium system, the frequency
dependent moduli are obtained from Fourier transforming:
\beq{d6} G'(\omega) + i\, G''(\omega)  =  i\, \omega \,
\int_0^\infty dt\; e^{-i\, \omega\, t}\;  v_\sigma \left.
\Phi^2(t)\right|_{\dot\gamma=0} + i \omega\, \eta_\infty\; . \eeq
Because of the vanishing of the Fourier-integral in \gl{d6} for high
frequencies, the parameter $\eta_\infty$ can be identified as high
frequency viscosity:
\beq{d7} \lim_{\omega\to\infty} G''(\omega) / \omega
=\eta^{\omega}_\infty\quad , \; \mbox{with }\;\eta^{\omega}_\infty =
\eta_{\infty}\; . \eeq
At high shear, on the other hand, \gl{d2} leads to a vanishing of
$m(t)$, and \gl{d1} gives an exponential decay of the transient
correlator, $\Phi(t) \to e^{-\Gamma\, t}$ for $\gd\to0$. The high
shear viscosity thus becomes \beq{d8} \eta^{\gd}_\infty =
\lim_{\gd\to\infty} \sigma(\gd)/ \gd = \eta_\infty +
\frac{v_\sigma}{2\, \Gamma} = \eta^\omega_\infty +
\frac{v_\sigma}{2\, \Gamma} \; . \eeq

Representative solutions of the F$_{12}^{(\gd)}$-model are
summarized in Fig.\ref{Figneu}, which bring out the discussed
universal aspects included in all ITT models.

\section{Experimental system and methods}

The particles consist of a solid core of poly(styrene) onto which a
network of crosslinked poly(N-isopropylacrylamide) (PNIPAM) is
affixed. The degree of crosslinking of the PNIPAM shell effected by
the crosslinker N,N'-methylenebisacrylamide (BIS)  was 5 Mol $\%$.
The core-shell type PS-NIPAM particles were synthesized, purified
and characterized as described in ref.\cite{Din:98} Immersed in
water the shell swells at low temperatures.
Raising the temperature above 32$^o$C leads to a volume transition
within the shell. To investigate the structure
and swelling of the particles cryogenic transmission electron microscopy and
dynamic light scattering have been used.

Screening the remaining electrostatic interactions by adding 5.10$^{-2}$ molL$^{-1}$ KCl, the system crystallises as hard
spheres.\cite{Cra:06} Experimental details on the characterization
of the particles and on the determination of the effective volume
are given elsewhere.\cite{Cra:07} The dependence of $\phi_{\rm eff}$
on the temperature is given by the hydrodynamic radius $R_H$
determined from the dynamic light scattering in the dilute regime.
$R_H$ was linearly extrapolated between 14 and 25$^o$C
($R_H=-0.85925T +123.78$ with $T$ the temperature in $^o$C) as
described recently\cite{Cra:06} and $\phi_{\rm eff}$ was calculated
following the relation
\begin{equation} \label{equ4}
\phi_{\rm eff} =k\; c \left(\frac{R_H}{R}\right)^3
\end{equation}
with $R$ the radius of the core determined by cryogenic transmission
electron microscopy ($R=52$ nm),\cite{Cra:06} $c$ the concentration
in wt $\%$ and $k$ a rescaling constant. In order to determine $k$,
an experimental phase diagram has been achieved by determining the
crystal fraction of the samples from the position of the coexistence
liquid-crystal boundaries after sedimentation. This was linearly
extrapolated to identify the beginning and the end of the
coexistence domain. The experimental phase diagram of the
suspensions of the core-shell particles  was rescaled with the
constant $k=0.4814$ to the freezing volume fraction for hard spheres
$\phi_F=0.494$.\cite{Hoo:71}

Three instruments were employed in the present study to investigate
the rheological properties of the suspensions. The flow behaviour
and the linear viscoelastic properties for the range of the low
frequencies were measured with a stress-controlled rotational
rheometer MCR 301 (Anton Paar), equipped with a Searle system (cup
diameter: 28.929 mm, bob diameter: 26.673 mm, bob length: 39.997
mm). Measurements have been performed on 12 ml solution and the
temperature was set with an accuracy of $\pm$ 0.05$^o$C. The shear
stress $\sigma$ versus the shear rate $\dot{\gamma}$ (flow curve)
was measured after a pre-shearing of $\dot{\gamma} = 100$ s$^{-1}$
for two minutes and a timesweep of 1 hour at 1 Hz and 1 $\%$
deformation in the linear regime. The flowcurves experiment were
performed setting $\dot{\gamma}$, first with increasing
$\dot{\gamma}$ from $\dot{\gamma} =10^{-4}-10^3$ s$^{-1}$ with a
logarithmic time ramp from 600 to 20 s, and then with decreasing
$\dot{\gamma}$. The stationarity has been checked by step-flow
experiments in the glassy state for the highest effective volume
fraction ($\phi_{\rm eff}=0.622)$. The frequency dependence of the
loss $G''$ and elastic $G'$ moduli has been measured for 1 $\%$
strain from 15 to 10$^{-3}$ Hz with a logarithmic time ramp from 20
to 600 s. The dependence upon the strain has been checked and
confirmes that all the measurements were performed in the linear
regime. The frequency dependence was tested for two different sample histories. The experiments were first performed without pre-shearing
after the timesweep, before the flowcurves experiments, and then
after the flowcurves experiments 10 s after two minutes pre-shearing
at $\dot{\gamma}= 100$ s$^{-1}$ to melt eventual crystallites. We
only considered experiments performed after pre-shearing in the
following discussion of $G'$ and $G''$ for the lowest frequencies.

Additional rheological experiments were carried out on Piezoelectric
vibrator (PAV)\cite{Cra:05} and cylindrical torsional
resonator\cite{Dei:01,Fri:03} supplied by the Institut f\"{u}r
dynamische Materialpr\"{u}fung, Ulm, Germany. The PAV was operated
from 10 to 3000 Hz. The solution is placed between two thick
stainless steel plates. The upper one remains static whereas the lower is
cemented to piezoelectric elements. The gap was adjusted with a 100
$\mu m$ ring. One set of piezoelectric elements is driven by an ac
voltage to induce the squeezing of the material between the two
plates, whereas the second set gives the output voltage. Experimental
details concerning this instrument are given elsewhere.\cite{Cra:05}
Only the measurements in the glassy state have been performed with
the PAV as the instrument does not allow any pre-shearing.

The cylindrical torsional resonator used was operated at a single
frequency (26 kHz). The experimental procedure and the evaluation of
data have been described recently.\cite{Dei:01,Fri:03}\\

The effect of the shear rate $\gd$ on the particle dynamics is
measured by the Peclet number,\cite{russel} Pe$_0=\gd R_H^2/D_0$,
which compares the rate of shear flow with the time an isolated
particle requires to diffuse a distance identical to its radius.
Similarly, frequency will be reported in the following rescaled by
this diffusion time, $\omega'= \omega  R_H^2/D_0$. The self
diffusion coefficient $D_0$ at infinite dilution was calculated from
the hydrodynamic radius $R_H$ and the viscosity of the solvent
$\eta_s$ with the Stokes-Einstein relation so that
$D_0=k_BT/6\pi\eta_S R_H$. In dense dispersions, however, the
structural rearrangements proceed far slower than diffusion at
infinite dilution, and therefore, very small Peclet numbers and
rescaled frequencies $\omega'$ are of interest in the following.
Stresses will be measured in units of $k_BT/R_H^3$ in the following.

\section{Comparison of theory and experiment}

Shear stresses measured in non-linear response of the dispersion
under strong steady shearing, and frequency dependent shear moduli
arising from thermal shear stress  fluctuations in the quiescent
dispersion were measured and fitted with results from the schematic
F$_{12}^{(\gd)}$-model. Some results from the microscopic MCT for
the equilibrium moduli are included also; see Sect.~VI.C for more
details.\cite{numerics} In the following discussion, we first start
with more general observations on typical fluid and glass like data,
and then proceed to a more detailed analysis. Figures \ref{Fig1} and
\ref{Fig2} show measurements in fluid states, at $\phi_{\rm
eff}=0.540$ and $\phi_{\rm eff}=0.567$, respectively, while
Fig.~\ref{Fig3} was obtained in the glass at $\phi_{\rm eff}=0.627$.
From the fits to all $\phi_{\rm eff}$, the glass transition value
$\phi_{\rm eff}^c=0.58$  was obtained, which agrees well with the
measurements on classical  hard sphere
colloids.\cite{Meg:93,Meg:94,Bar:02}

\subsection{Crystallisation effects}

We start the comparison of experimental and theoretical results by
recalling the interpretation of time in the ITT approach. Outside
the linear response regime,  both $\Phi(t)$ and $g(t)$ describe the
decorrelation  of  equilibrium, fluid-like fluctuations under shear
and internal motion. Integrating through the transients provides the
steady state averages, like the stress. While theory finds that
transient fluctuations always relax under shear, real systems may
either remain in metastable states if $\dot\gamma$ is too small to
shear melt them, or undergo transitions to heterogeneous states for
some parameters. In these circumstances, the theory can not be
applied, and the rheological response of the system, presumably, is
dominated by the  heterogeneities. Thus, care needs to be taken in
experiments, in order to prevent phase transitions, and to shear
melt arrested structures, before data can be recorded.

\begin{figure}[t]
\centering
\includegraphics[ width=0.8\columnwidth]{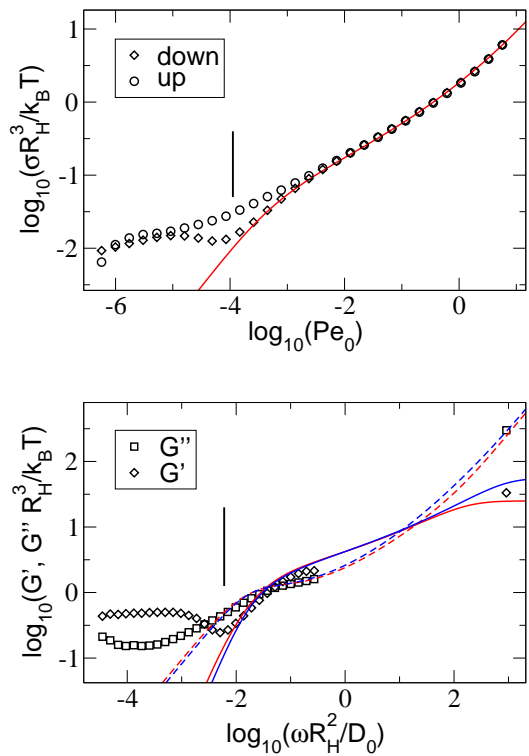}
\bigskip
\caption{ The reduced flow curves and the corresponding moduli for a
fluid state at 13.01wt\%, T$=20^oC$, and $\phi_{\rm eff}=0.540$.
Flow curves measured proceeding from higher to lower shear rates
(called 'down' flow curves) and dynamic experiments were fitted
where effects from crystallisation can be neglected; the lower
limits of the unaffected-data regions are marked by vertical bars.
The red lines show the fits with the schematic
$F_{12}^{\left(\dot{\gamma}\right)}$-model while the blue lines show
the results from microscopic MCT (solid $G'$, broken $G''$), with
parameters: $\epsilon=-0.05$, $\frac{D_{S}}{D_{0}}=0.15$, and
$\eta_{\infty}=0.3\, k_{B}T/(D_{0}R_{H}) $; the moduli were scaled
up by a factor $c_{y}=1.4$. \label{Fig1} }
\end{figure}

\begin{figure}[t]
\centering
\includegraphics[ width=0.8\columnwidth]{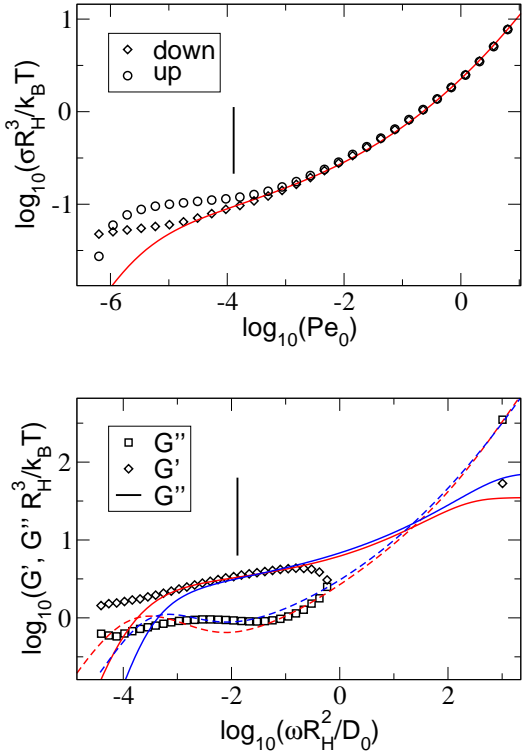}
\bigskip
\caption{The reduced flow curves and the corresponding moduli (like in Fig. \ref{Fig1}) for a
fluid state  at 13.01wt\%, T$=18^oC$, and $\phi_{\rm eff}=0.567$. The
vertical bars mark the minimal Peclet number or rescaled frequency
for which the influence of crystallisation can be neglected.
Microscopic parameters: $\epsilon=-0.01$,
$\frac{D_{S}}{D_{0}}=0.15$, and $\eta_{\infty}=0.3\,
k_{B}T/(D_{0}R_{H}) $; moduli scale factor $c_{y}=1.4 $.
\label{Fig2} }
\end{figure}

The small size polydispersity of the present particles enables the
system to grow crystallites according to its equilibrium phase
diagram. Fortunately, when recording flow curves, viz. stress as
function of shear rate, data can be taken when decreasing the shear
rate. We find that the resulting 'down' flow curves correspond to
amorphous states and reach the expected low-$\gd$ asymptotes
($\sigma=\eta_0\,\gd$, see Fig.~\ref{Fig1}), except for very low
$\gd$, when an increase in stress indicates the formation of
crystallites. 'Up' flow curves, however, obtained when moving
upwards in shear rate during the measurement of the stress are
affected by crystallites formed after the initial shearing at 100
s$^{-1}$ during the timesweep and the first frequency sweep
experiments. See the hysteresis between 'up' and
'down' flow curves in Figs.~\ref{Fig1} and \ref{Fig2}, where
measurements for two fluid densities are reported. Above a critical 
shear rate $\dot{\gamma}_{\rm cr}\approx
4s^{-1}$ no hysteresis has been observed, which proved that all the
crystallites have been molten.  In the present
work we focus on the 'down' flow curves, and consider only data
taken either for $\dot{\gamma}>\dot{\gamma}_{\rm cr}$, or (for
$\dot{\gamma}<\dot{\gamma}_{\rm cr}$)  before the time crystallisation sets in.
This time was estimated from
timesweep experiments as the time where crystallisation caused a 10$\%$ deviation of the complex modulus $G^*$. Thus, only the portions of the flow
curves unaffected by crystallisation are taken into account; in Figs.~\ref{Fig1} and \ref{Fig2} vertical bars denote the limits. We
find that the effect of crystallisation  on the flow curves is
maximal around $\phi=0.55$ and becomes progressively smaller and
shifts to lower shear rates for higher densities; see
Figs.~\ref{Fig1}, \ref{Fig2} and \ref{Fig3}. This agrees with the
notion that the glass transition slows down the kinetics of
crystallisation and causes the average size of crystallites to
shrink.\cite{Meg:94} For the highest densities, which are in the
glass without shear, the hysteresis at the lowest $\gd$ can be
attributed to a non stationarity of the up curve; see
Fig.~\ref{Fig3}. This effect has been confirmed by step flow
experiments, but does not affect the back curves.

The linear response moduli similarly are   affected by the presence
of small crystallites at low frequencies. $G'(\omega)$ and
$G''(\omega)$ increase above the behavior expected for a solution
($G'(\omega\to0)\to\eta_0\, \omega$ and $G''(\omega\to0)\to c\,
\omega^2$) even at low density, and exhibit elastic contributions at
low frequencies (apparent from $G'(\omega)>G''(\omega)$); see
Figs.~\ref{Fig1} and \ref{Fig2}.  This effect follows the
crystallisation of the system during the measurement after the
shearing at $\dot{\gamma}=100$ $s^{-1}$. Only data will be
considered in the following 
which were collected before the crystallisation time. For higher effective
volume fraction other effects such as ageing and an ultra-slow process had to be
taken into account and will be discussed more in detail in the next
section.

\subsection{Shapes of flow curves and moduli and their relations}

\begin{figure}[t]
\centering
\includegraphics[ width=0.8\columnwidth]{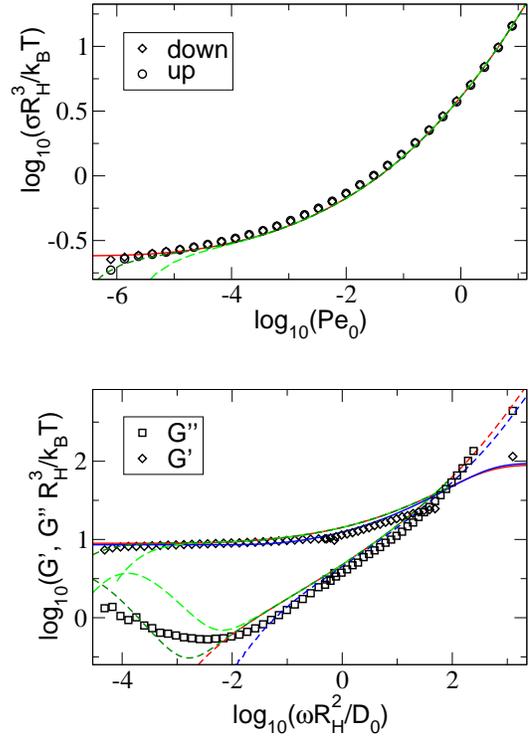}
\bigskip
\caption{The reduced flow curves and the corresponding moduli for a
glass state at 13.01wt\%, T$=14^oC$, and $\phi_{\rm eff}=0.622$. See
figure \ref{Fig1}  for further explanations.  Microscopic
parameters: $\epsilon=0.03$,  $\frac{D_{S}}{D_{0}}=0.08$,  and
$\eta_{\infty}=0.3\,  k_{B}T/(D_{0}R_{H}) $; moduli scale factor
$c_{y}=1.4 $ (blue). Curves from the schematic
F$_{12}^{(\gd)}$-model with an additional dissipative process
included (Eq.~\ref{ap1}) are shown as dashed lines; $\delta=
10^{-7}\, \Gamma$ (long dashes, light green) and $\delta= 10^{-8}\,
\Gamma$ (short dashes, dark green). Here $\Gamma=88\, D_0/R_H^2$.
The red curves give the schematic model calculations for identical
parameters but without additional dissipative process
(viz.~$\delta=0$). \label{Fig3} }
\end{figure}

The flow curves and moduli exhibit a qualitative change when
increasing the effective packing fraction from around 50\% to above
60\%. For lower densities (see Fig.~\ref{Fig1}), the flow curves
exhibit a Newtonian viscosity $\eta_0$ for small shear rates,
followed by a sublinear increase of the stress with $\gd$; viz.~a
region of shear thinning behavior. For the same densities, the
frequency dependent spectra exhibit a broad peak or shoulder, which
corresponds to the final or $\alpha$-relaxation discussed in
Sect.~III. Its peak position (or alternatively the crossing of the
moduli, $G'=G''$) is roughly given by $\omega \tau=1$ (see
Fig.~\ref{Fig2}). These properties characterize a viscoelastic
fluid. For higher density, see Fig.~\ref{Fig3}, the stress in the
flow curve remains above a finite yield value even for the smallest
shear rates investigated. The corresponding storage modulus exhibits
an elastic plateau at low frequencies. The loss modulus drops far
below the elastic one. These observations characterize a soft solid.
The loss modulus rises again at very low frequencies, which may
indicate that the colloidal solid at this density is metastable and
may have a finite lifetime (an ultra-slow process is discussed in
Sect.~VI.E).

Simple relations, like the 'Cox-Merz rule', have sometimes been used
in the past to compare the shapes of the flow curves $\sigma(\gd)$
with the shapes of the dissipative modulus $G''(\omega)$. Both
quantities can be interpreted in terms of a (generalized) viscosity,
on the one hand as function of shear rate
$\eta(\gd)=\sigma(\gd)/\gd$, and on the other hand as function of
frequency $\eta(\omega)=G''(\omega)/\omega$. The Cox-Merz rule
states that the functional forms of both viscosities coincide.

Figures \ref{Fig1} to \ref{Fig3} provide a sensitive test of
relations in the shapes of $\sigma(\gd)$ and $G''(\omega)$. Figure
\ref{Fig2} shows most conclusively, that no simple relation between
the far-from equilibrium stress as function of external rate of
 shearing exists with the equilibrium  stress fluctuations
at the corresponding frequency. While $\sigma(\gd)$ increases
monotonically, the dissipative modulus $G''(\omega)$ exhibits a
minimum for fluid states close to the glass transition. It separates
the low-lying final relaxation process in the fluid from the
higher-frequency relaxation.

As shown in Fig.~\ref{Figneu}, the frequency dependence of $G''$ in
the minimum region is given by the scaling function $\cal G$ of
Sect.~III, which describes the minimum as crossover between two
power laws. The approximation for the modulus around the minimum
\beq{e1} G''(\omega) \approx \frac{G_{\rm min}}{a+b} \; \left[\, b\,
\left( \frac{\omega}{\omega_{\rm min}}\right)^a + a\, \left(
\frac{\omega_{\rm min}}{\omega}\right)^{b} \, \right] \eeq has been
found in the quiescent fluid ($\epsilon<0$, $\gd=0$), and works
quantitatively if the relaxation time $\tau$ is large, viz.~ time
scale separation holds  for small $|\epsilon|$.\cite{Goe:91} The
parameters in this approximation follow from \gls{c3}{c4} which give
$G_{\rm min}\propto\sqrt{-\epsilon}$ and $\omega_{\rm min}\propto
(-\epsilon)^{1/2a}$. For packing fractions too far below the glass
transition, the final relaxation process is not clearly separated
from the high frequency relaxation. This holds in Fig.~\ref{Fig1},
where the final structural decay process only forms a shoulder.
Closer to the transition, in Fig.~\ref{Fig2}, it is separated, but
crystallisation effects prevent us from fitting \gl{e1} to the data.

Asymptotic power-law expansions of $\sigma(\gd)$ exist close to the
glass transition, which were deduced from the stability analysis in
Sect.~III;\cite{Fuc:03,Hen:05,Haj:07} yet we refrain from entering
their detailed discussion and describe the qualitative behavior in
the following. For the same parameters in the fluid, where the
minimum in $G''(\omega)$ appears, the flow curves follow a S-shape
in a double logarithmic plot, crossing over from a linear behavior
$\sigma=\eta_0\, \gd$ at low shear rates to a downward curved piece,
followed by a point of inflection, and an upward curved piece, which
finally goes over into a second linear behavior at very large shear
rates, where $\sigma=\eta^{\gd}_\infty\, \gd$. This S-shape can be
recognized in Figs.~\ref{Fig1} and \ref{Fig2}. Because of the finite
slope of $\log_{10}{\sigma}$ versus $\log_{10}{\gd}$ at the point of
inflection, one may speculate about an effective power-law
$\log_{10}{\sigma} \approx c + c' \log_{10}{\gd}$. In
Fig.~\ref{Fig1} this happens at Pe$_0\approx 10^{-2}$. Yet, the
power-law is only apparent because the point of inflection moves,
the slope changes with distance to the glass transition, and the
linear bit in the flow curve never extends over an appreciable
window in $\gd$.\cite{Haj:07}

\begin{figure}[t]
\centering
\includegraphics[ width=0.8\columnwidth]{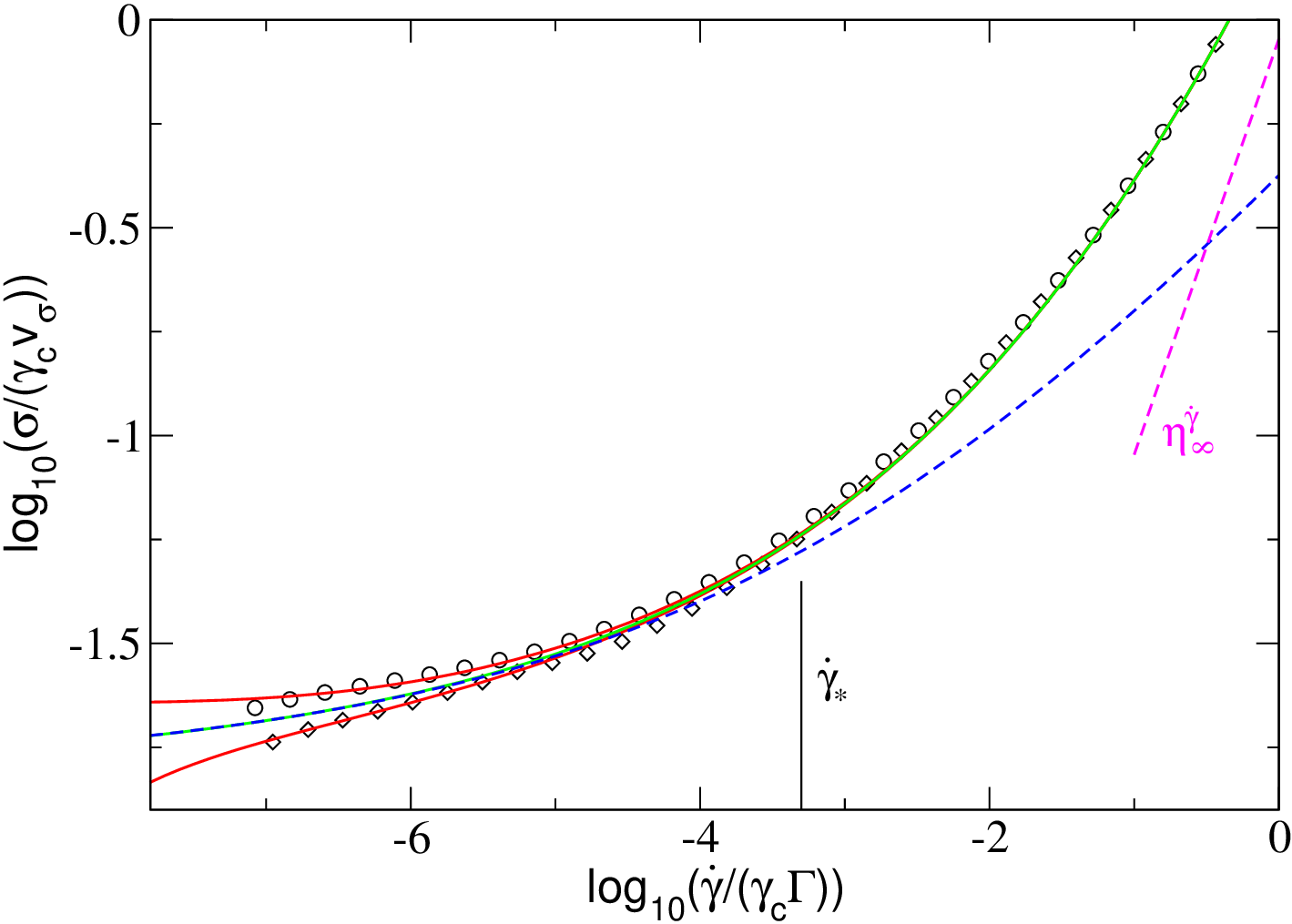}
\bigskip
\caption{Flow curves for $\phi_{\rm eff}=0.580$ (diamonds) and
$\phi_{\rm eff}=0.608$ (circles) rescaled with the parameters from
the respective F$_{12}^{(\gd)}$-model fits; the corresponding fits
are given as red solid lines, with separation parameter
$\epsilon=-2\, 10^{-4}$ ($\phi_{\rm eff}=0.580$) and $\epsilon=3\,
10^{-4}$ ($\phi_{\rm eff}=0.608$). The blue dashed cuves gives the
generalized Hershel-Bulkley law from \gl{e2}, which holds at
$\epsilon=0$;  for the F$_{12}^{(\gd)}$-model, its parameters are
$\gd_*= 5.06\, 10^{-5} \Gamma= 2.16\, 10^{-5} / t_0$ (corresponding
to a Peclet number Pe$_0^*=4.25\, 10^{-3}$ at $\phi_{\rm
eff}=0.608$), $c_2\approx 0.936$, and $c_3\approx 0.632$. The green
solid curve is the (critical) flow curve at  $\epsilon=0$ . The
magenta dashed straight line labeled $\eta_\infty^{\gd}$ denotes the
high-shear asymptote \gl{d8}. \label{Fighb} }
\end{figure}

Non-trivial power-laws in the flow curves exist close to the
transition itself. At $\epsilon=0$, a generalized Hershel-Bulkley
law holds
\beq{e2} \sigma(\gd\to0) \to \sigma^+_c \left( 1 +
|\frac{\gd}{\gd_*}|^m + c_2 \, |\frac{\gd}{\gd_*}|^{2m} + c_3 \,
|\frac{\gd}{\gd_*}|^{3m }  \right) \eeq
and describes the flow curves over an appreciable part of the range
$\gd\le\gd_*$, where structural relaxation dominates the
stress;\cite{Hen:05,Haj:07} the exponent is $m=0.143$ in the
F$_{12}^{(\gd)}$-model for this $\lambda$. It provides a
semi-quantitative fit  of the flow curves for more than a decade in
$\gd$ close to the glass transition as shown in Fig.~\ref{Fighb}.
There $|\epsilon|$ is quite small at these effective packing
fractions.  A qualitative difference of the glass flow curves to the
fluid S-shaped ones, is that the shape of $\sigma(\gd)$ constantly
has an upward curvature in double-logarithmic representation.  The
yield stress can be read off by extrapolating the flow curve to
vanishing shear rate. In Fig.~\ref{Fig3} this leads to a value
$\sigma^+\approx 0.24\, k_BT/R_H^3$ at $\phi_{\rm eff}=0.622$, which
is in agreement with previous measurements in this system over a
much reduced window of shear rates.\cite{Fuc:05a,Cra:06} While this
agreement supports the prediction of an dynamic yield stress in the
ITT approach, and demonstrates the usefulness of this concept, small
deviations in the flow curve at low $\gd$ are present in
Fig.~\ref{Fig3}.  We postpone to Sect.~VI.E the discussion of these
deviations, which indicate the existence of an additional slow
dissipative process in the glass. Its signature is seen most
prominently in the loss modulus $G''(\omega)$ in Fig.~\ref{Fig3}.

The storage modulus of the glass shows striking elastic behavior.
$G'(\omega)$ exhibits a near plateau over more than three decades in
frequency, which allows to read off the elastic constant $G_\infty$
easily.

\subsection{Microscopic MCT results}

Included in figures \ref{Fig1} to \ref{Fig3} are calculations using
the microscopic MCT given by \glto{b7}{b11} evaluated for hard
spheres.\cite{numerics} This is presently possible without shear
only ($\gd=0$), because of the complications arising from anisotropy
and time dependence in \gl{b9}. The  a priori unknown,
adjustable parameter is the matching time scale $t_0$, which we
adjusted by varying the short time diffusion coefficient appearing
in the initial decay rate in \gl{b8}. The computations were
performed with $\Gamma_{\qb}(t)\equiv\Gamma_q=D_s\, q^2/S_q$, and
values for $D_s/D_0$ are reported in the captions of
Figs.~\ref{Fig1} to \ref{Fig3}, and in table I. The viscous
contribution to the stress is again mimicked by including
$\eta_\infty$ like in \gl{d6}.

Gratifyingly, the stress values  computed from the microscopic
approach are close to the measured ones; they are too small by 40\%
only, which may arise from the approximate structure factors
entering the MCT calculation; the Percus-Yevick approximation was
used here.\cite{russel} In order to compare the shapes of the moduli
the MCT calculations were scaled up by a factor $c_y=1.4$ in
Figs.~\ref{Fig1} to \ref{Fig3}. Microscopic MCT also does not hit
the correct value for the glass transition
point.\cite{Goe:92,Goe:91} It finds $\phi^{\rm MCT}_c=0.516$, while
our experiments give $\phi^{\rm exp}_c\approx0.58$. Thus, when
comparing, the relative separation from the respective transition
point needs to be adjusted as, obviously, the spectra depend
sensitively on the distance to the glass transition; the fitted
values of the separation parameter $\epsilon$ are included in
Fig.\ref{Fig5}.

\begin{figure}[t]
\centering
\includegraphics[ width=0.8\columnwidth]{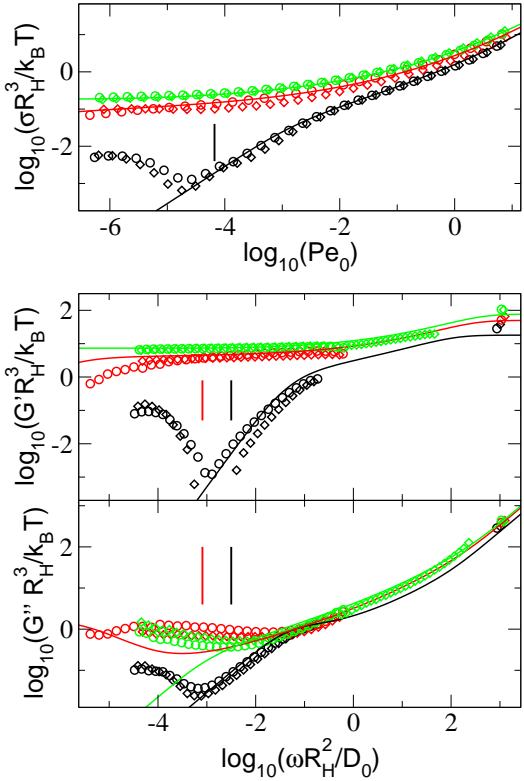}
\bigskip
\caption{
 The plots demonstrate that the reduced flow curves and the reduced moduli
 are unique functions only depending on
$\phi_{\rm eff}$.
 All flow curves are down curves.
 The fits using the schematic F$_{12}^{(\gd)}$-model were performed with the data
 points at 13.01wt\% taken before the onset of crystallisation (data to the right of  the vertical bars).
 Black diamonds: 12.10wt\% and
$\phi_{\rm eff}=0.527$.
 Black circles: 13.01wt\% and
$\phi_{\rm eff}=0.527$.
 Red diamonds: 12.10wt\% and $\phi_{\rm eff}=0.578$.
 Red circles: 13.01wt\% and $\phi_{\rm eff}=0.580$.
 Green diamonds: 13.01wt\% and $\phi_{\rm eff}=0.608$.
 Green circles: 13.58wt\% and $\phi_{\rm eff}=0.606$.
\label{Fig4} }
\end{figure}

Considering the low frequency spectra in $G'(\omega)$ and
$G''(\omega)$, microscopic MCT and schematic model provide
completely equivalent descriptions of the measured data. Differences
in the fits in Figs.~\ref{Fig1} to \ref{Fig3} for $\omega
R_H^2/D_0\le 1$ only remain because of slightly different choices of
the fit parameters which were not tuned to be close. These
differences serve to provide some estimate of uncertainties in the
fitting procedures. Main conclusion of the comparisons is the
agreement of the moduli from  microscopic MCT, schematic ITT model,
and from the measurements. This observation strongly supports the
universality of the glass transition scenario which is a central
line of reasoning in the ITT approach to the non-linear rheology.

At large $\gd$ and large $\omega$ hydrodynamic interactions become
important. In the flow curves, $\eta_\infty^{\gd}$, and, in the loss
modulus, $\eta_\infty^{\omega}$ become relevant parameters, and the
structural relaxation captured in ITT and MCT is not sufficient
alone to describe the rheology. Qualitative differences appear in
the moduli, especially in $G'(\omega)$, between the schematic model
and the microscopic MCT. While the storage modulus of the
F$_{12}^{(\gd)}$-model crosses over to a high-$\omega$ plateau
already at rather low $\omega$, the microscopic modulus continues to
increase for increasing frequency, especially at lower densities;
see the region $\omega \,\gtrapprox\, 10^2 D_0/R_H^2$ in
Figs.~\ref{Fig1} to \ref{Fig2}. The latter aspect is connected to
the high-frequency divergence of the shear modulus of particles with
hard sphere potential,\cite{Mas:95} as captured within the MCT
approximation.\cite{Nae:98,numerics}   As carefully discussed by
Lionberger and Russel, lubrication forces may suppress this
divergence and its observation thus depends on the surface
properties of the colloidal particles.\cite{Lio:94} Clearly, the
region of (rather) universal properties arising from the
non-equilibrium transition between shear-thinning fluid and yielding
glass is left here, and particle specific effects become important.

\subsection{Parameters}

In the microscopic ITT approach from Sect.~II, the rheology is
determined from the equilibrium structure factor $S_q$ alone. This
holds at low enough frequencies and shear rates, and excludes the
time scale parameter $t_0$ of \gl{c3}, which needs to be found by
matching to the short time dynamics. This prediction has as
consequence that the flow curves and moduli should be a function
only of the thermodynamic parameters characterizing the present
system, viz.~its structure factor.

Figure \ref{Fig4} supports this claim by proving that the
rheological properties of the dispersion only depend on the
effective packing fraction, if particle size is taken account of
properly. Figure \ref{Fig4} collects flow curves and moduli measured
for different concentrations of particles according to weight, and
for different radii $R_H$ adjusted by temperature. Whenever the
effective packing fraction, $\phi_{\rm eff}=(4\pi/3) n R_H^3$, is
close, the  rheological data overlap in the window of structural
dynamics. Obviously, appropriate scales for frequency, shear rate
and stress magnitudes need to be chosen to observe this. The
dependence of the vertices on $S_q$ (\gls{b7}{b10}) suggests that
$k_BT$ sets the energy scale as long as repulsive interactions
dominate the local packing. The length scale is set by the average
particle separation, which can be taken to scale with $R_H$ in the
present system. The time scale of the glassy rheology within ITT is
given by $t_0$ from \gl{c3}, which we take to scale with the
measured dilute diffusion coefficient $D_0$. Thus the rescaling of
the rheological data can be done with measured parameters alone.
Figure \ref{Fig4} shows quite satisfactory scaling. Whether the
particles are truely hard spheres is not of central importance to
the data collapse in Fig.~\ref{Fig4} as long as the static structure
factor agrees for the $\phi_{\rm eff}$ used. Fits with the
F$_{12}^{(\gd)}$-model to all data are possible, and are of
comparable quality to the fits shown in Figs.~\ref{Fig1} to
\ref{Fig3}.

\begin{figure}[t]
\centering
\includegraphics[ width=0.8\columnwidth]{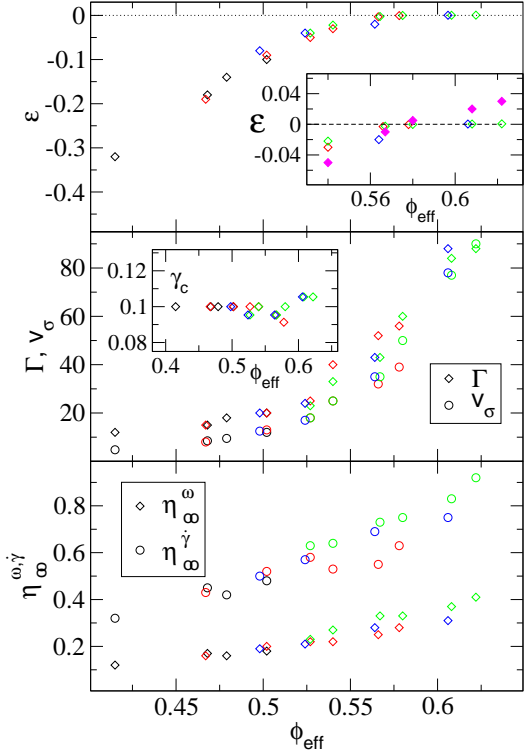}
\caption{ The fitted parameters of the F$_{12}^{(\gd)}$-model (open
symbols). Black symbols: 10.75wt\%,  red symbols: 12.10wt\%, green
symbols: 13.01wt\%,  blue symbols: 13.58 wt\%. $\epsilon$ and
$\gamma_{c}$  are dimensionless. Filled magenta symbols, included in
the upper inset, give the $\epsilon$ values fitted in the
microscopic MCT calculations for 13.01wt\%. The unit of $v_{\sigma}$
is $k_{B}T/R_{H}^{3}$  while $\Gamma$  is given in units of
$D_{0}/R_{H}^{2}$. The high frequency and high shear viscosities
$\eta_{\infty}^{\omega,\dot{\gamma}}$ are given in units of
$k_{B}T/(D_{0}R_{H})$. \label{Fig5}}
\end{figure}

The fitted parameters used in the schematic F$_{12}^{(\gd)}$-model
are summarized in Fig.~\ref{Fig5}. Parameters corresponding to
identical concentrations by weight are marked by identical colours.
Within the scatter of the data one may conclude that all fit
parameters depend on the effective packing fraction only. This again
supports the mentioned dependence of the glassy rheology on the
equilibrium structure factor. The initial rate $\Gamma$, which sets
$t_0$,  appears a unique function of $\phi_{\rm eff}$, also; an
observation which is not covered by the present ITT approach. It
suggests that hydrodynamic interactions appear determined by
$\phi_{\rm eff}$ in the present system also.

Importantly, all fit parameters exhibit smooth and monotonous drifts
as function of the external thermodynamic control parameter,
viz.~$\phi_{\rm eff}$ here. Nevertheless,  the moduli at low
frequencies (e.g.~$G'(\omega)$ at $\omega R_H^2/D_0=0.01$), or the
stresses at low shear rates (e.g.~$\sigma(\gd)$ at $\gd
R_H^2/D_0=10^{-4}$) change by more than an order in magnitude in
Figs.~\ref{Fig1} to \ref{Fig3}.  Even larger changes may be obtained
from taking experimental data not shown, whose fit parameters are
included in Fig.~\ref{Fig5}. It is this sensitive dependence of the
rheology on small changes of the external control parameters that
ITT addresses.

\begin{figure}[t]
\centering
\includegraphics[width=0.8\columnwidth]{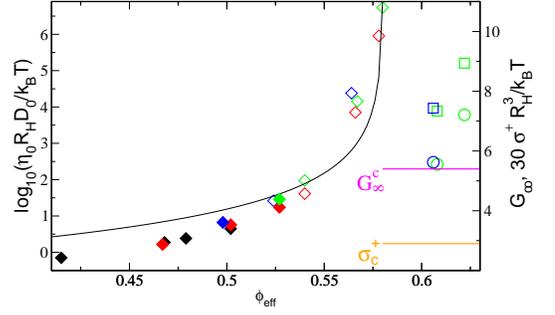}
\caption{ Newtonian viscosity $\eta_0$ (diamonds, left axis),
elastic constant $G_\infty$ (squares), and yield stress $\sigma^+$
(circles; data rescaled by a factor 30; both $G_\infty$ and
$\sigma^+$ right axis),  as functions of the effective packing
fraction $\phi_{\rm eff}$ as obtained from the fits performed with
the  F$_{12}^{(\gd)}$-model. Filled symbols indicate where direct
measurements of $\eta_0$  were possible. Black symbols: 10.75wt\%,
red symbols: 12.10wt\%, green symbols: 13.01wt\%, blue symbols:
13.58 wt\%. The line gives a power-law fit to the viscosity-data
over the full range using the known $\gamma=2.34$ exponent from MCT,
$\log_{10}{\eta_0}=A-\gamma\cdot\log_{10}\left(\phi_{\rm
eff}^c-\phi_{\rm eff}\right)$; the critical packing fraction is
found as $\phi_{\rm eff}^c=0.580$. Horizontal bars denote the
critical elastic constant $G_\infty^c$, and the critical yield
stress $\sigma^+_c$ as extrapolated from Fig.~\ref{Fighb}.
\label{Fig6}}
\end{figure}

When comparing the parameters from the schematic model to the ones
obtained from the microscopic MCT calculation of the moduli, one
observes qualitative and semi-quantitative agreement; see the
captions to Figs.~\ref{Fig1} to \ref{Fig3}, table I, and the upper
inset of Fig.~\ref{Fig5}. For example, the increase of the prefactor
$v_\sigma$ of stress fluctuations is captured in the microscopic
vertex where $S_q$ enters (this follows because the rescaling factor
$c_y$ is density independent). Also the hydrodynamic viscosity
$\eta_\infty=\eta_\infty^\omega$ roughly agrees and may be taken
$\phi_{\rm eff}$-independent in the fits with the microscopic
moduli. The fitted values of $\eta_\infty$ are actually not too
different from data obtained in Stokesian dynamics simulations of
true hard spheres, supporting our simplified view on the particle
interactions.\cite{Fos:00} On closer inspection, one may notice
that the separation parameter of the microscopic hard sphere
calculation obtains larger positive values than $\epsilon$ fitted
with the schematic model. Moreover, it follows an almost linear
dependence on the effective packing fraction as asymptotically
predicted by MCT, $\epsilon\approx 0.65\, (\phi_{\rm
eff}-\phi^c_{\rm eff})/\phi^c_{\rm eff}$ with glass transition
density $\phi_c=0.587$ slightly higher than from the schematic model
fits. The differing behavior of the separation parameter from the
fits with the F$_{12}^{(\gd)}$-model in the glass is not understood
presently. The microscopic calculation signals glassy arrest more
clearly than the schematic model fit. The short time diffusion
coefficient $D_s/D_0$ in the microscopic calculation decreases as
expected from considerations of hydrodynamic interactions.
Gratifyingly we find values in the range of
the short-time self diffusion coefficient observed in
Stokesian dynamics simulations for hard spheres.\cite{Ban:03} The
initial rate $\Gamma$, however, of the schematic model increases
with packing fraction. The ad hoc interpretatation of $\Gamma$ as
microscopic initial decay rate evaluated for some typical wavevector
$q_*$, viz.~the ansatz $\Gamma=D_s q_*^2/S_{q_*}$, thus apparently
does not hold.

\begin{table}\centering
\begin{tabular}[b]{|c||c|c||c|c|}
\hline
$\phi_{\rm eff}$ & $\epsilon$ & $D_s/D_0$ & $\epsilon'$& $D'_s/D_0$ \\
\hline
0.527 & - 0.08 & 0.15 &  & \\
\hline
0.540 & - 0.05 & 0.15 & & \\
\hline
0.567 & - 0.01 & 0.15 & & \\
\hline
0.580 & 0.005 & 0.13 & - 0.01 & 0.15 \\
\hline
0.608 & 0.02 & 0.11 & - 0.003 & 0.15 \\
\hline
0.622 & 0.03 & 0.08 & - 0.003 & 0.15 \\
\hline
\end{tabular}
\caption{Parameters of the fits with the microscopic MCT to the
linear-response moduli $G'(\omega)$ and $G''(\omega)$. The first two
columns of separation parameter $\epsilon$ and short-time diffusion
coefficient ratio $D_s/D_0$ correspond to the fits shown in
Figs.~\ref{Fig1} to \ref{Fig3} and Fig.~\ref{Figadd} (solid lines),
while the second columns of $\epsilon'$ and $D'_s/D_0$ correspond to
the dashed-lines in Fig.~\ref{Figadd}; when no value is given, the
values from the first two columns apply. In all cases $c_y=1.4$ and
$\eta_\infty=0.3\, k_{B}T/(D_{0}R_{H})$ are used.}
\end{table}

While the model parameters adjusted in the fitting procedure only
drift smoothly with density, the rheological properties of the
dispersion change dramatically. Figure \ref{Fig6} shows the
Newtonian viscosity as obtained from extrapolations of the fits in
the F$_{12}^{(\gd)}$-model. It changes by 6 orders in magnitude.
From the combination of $G''(\omega)$- and flow curve data we can
follow this divergence over more than one decade in direct
measurement. From the divergence of $\eta_0$ the estimate of the
critical packing fraction can be obtained using the power-law
\gl{c4}, because the exponent $\gamma$ is known. We find $\phi_{\rm
eff}^c=0.580$ in nice agreement with the value expected for
colloidal hard spheres. On the glass side, the elastic constant and
yield stress jump discontinuously into existence. Reasonable values
are obtained from the F$_{12}^{(\gd)}$-model fits compared to data
from comparable systems. The strong increase of the elastic
quantities upon small increases of the density is apparent.

\subsection{Additional dissipative process in glass}

\begin{figure}[t]
\centering
\includegraphics[ width=0.8\columnwidth]{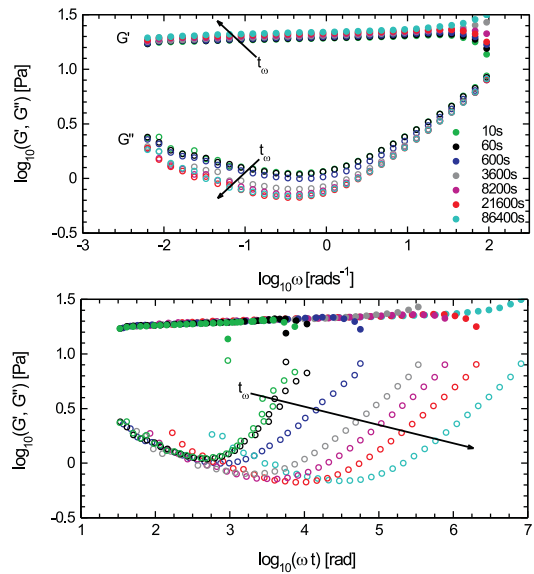}
\caption{The storage $G'(\omega)$ and loss $G''(\omega)$ moduli for
different waiting times $t_w$. The data have been also plotted as
function of $\omega t$ as suggested recently.\cite{Pur:06} See text
for further explanation.\label{Figageing} }
\end{figure}

One of the major predictions of the ITT approach concerns the
existence of glass states, which exhibit an elastic response for low
frequencies under quiescent conditions, and which flow only because
of shear and exhibit a dynamic yield stress under stationary shear.
Figure \ref{Fig3} shows such glassy behavior, as is revealed by the
analysis using the schematic and the microscopic model.
Nevertheless, the loss modulus $G''(\omega)$ rises at low
frequencies, clearly indicating the presence of a dissipative
process. It is not accounted for by the present theory. Also, the
storage modulus $G'(\omega)$ shows some downward bend at the lowest
frequencies.

These deviations can not be rationalized by ageing effects or
non-linearities in the response; see Fig.~\ref{Figageing}. We
checked the dependence on time since quench to this glass state and
also the linear dependence of the stress on the shear amplitude.
While we find ageing effects shortly after cessation of
pre-shear,\cite{Pur:06,Sol:97,Sol:98} these saturate after one day,
when the drifts of the spectra come to a stop. Ageing effects do not
change the spectra qualitatively, as the dissipative process appears
to possess a finite equilibrium relaxation time. As suggested
recently for dense PNIPAM microgel dispersions\cite{Pur:06} the same
data have been plotted as function of $\omega t$. Here, $t$ is the
total waiting time and is defined as function of the waiting time
$t_w$ before starting the measurement and the time
$\delta(t(\omega_n))$ expired between $t_w$ and the acquisition of
the data as $t = t(\omega_n) = t_w +\delta(t(\omega_n))$. The curves
collapse on a master curve in the low frequency range up to $\omega
t\approx 3000$ as expected from ageing theory for waiting time
$t_w<8200$ s. This prediction is no more respected for longer
waiting times, where an additional relaxation process is identified.
It cuts off the ageing behavior, when the age of the sample
approaches the value of its relaxation time. This supports the
introduction of an hopping phenomenon in our model, with a
characteristic relaxation time of the order of $10^4$ s ($\approx
R_H^2/\delta D_0=10^{8}R_H^2/\Gamma D_0=8.8 \, 10^{3}$ s see
Fig.\ref{Fig3}).

\begin{figure}[t]
\centering
\includegraphics[ width=0.8\columnwidth]{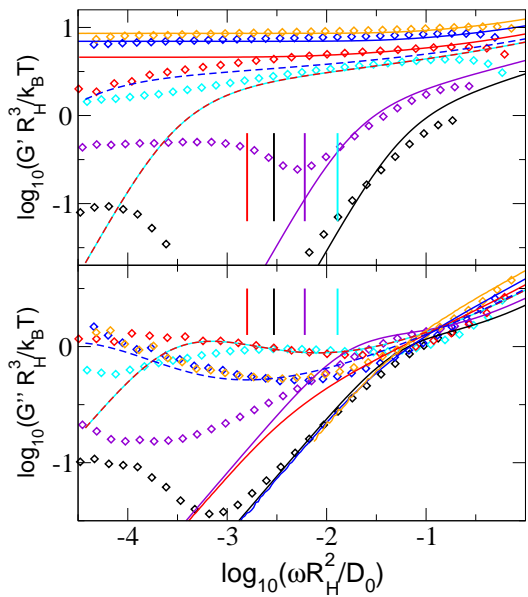}
\caption{Fits with microscopic MCT to the linear-response moduli
$G'(\omega)$ (upper panel) and $G''(\omega)$ (lower panel) for the
packing fractions $\phi_{\rm eff}=0.527$ (black diamonds and lines),
$\phi_{\rm eff}=0.540$ (violet), $\phi_{\rm eff}=0.567$ (light
blue), $\phi_{\rm eff}=0.580$ (red), $\phi_{\rm eff}=0.608$ (dark
blue), and $\phi_{\rm eff}=0.622$ (orange). Continuous lines give
the fits optimized for describing the storage modulus $G'(\omega)$;
these fits are also shown in Figs.~\ref{Fig1} to \ref{Fig3}, and the
corresponding parameters are included in Fig.~\ref{Fig5}, and
summarized in the left two columns in Table I. Broken lines for
$\phi_{\rm eff}=0.580$ (red, overlapping with the solid light blue
curve), and  $\phi_{\rm eff}=0.608$ (dark blue; the same curve would
fit $\phi_{\rm eff}=0.622$) show microscopic MCT calculations
attempting to fit the minima in $G''(\omega)$ enforcing negative
separation parameters $\epsilon$ (parameters included in table I).
These fluid like spectra can rationalize $G''(\omega)$, but fail
qualitatively to describe $G'(\omega)$. Vertical bars in corresponding colours denote the frequencies below which crystallisation affects the data at the different $\phi_{\rm eff}$. \label{Figadd} }
\end{figure}

Let us stress, moreover, that the state shown in Fig.~\ref{Fig3} is
not a fluid state within the present approach. The presence of an
elastic window in $G'(\omega)$, its increase as function of packing
fraction, and the upward curvature of the flow curves rule out a
negative separation parameter $\epsilon<0$ of this state at
$\phi_{\rm eff}=0.622$. Calculations within the microscopic MCT
document this convincingly. Figure \ref{Figadd} compares the MCT
calculations for hard spheres  with moduli ranging from fluid to
glassy states. By adjusting the effective packing fraction,  MCT
semi-quantitatively describes the dominating modulus, either loss or
storage one, for all states (corresponding curves already shown in
Figs~\ref{Fig1} to \ref{Fig3}). At high concentrations, it describes
the storage modulus $G'(\omega)$ on an error level of $1
k_BT/R_H^3$, and misses the loss modulus $G''(\omega)$ by a similar
absolute error. Yet, because the latter is itself of the order of
$G''(\omega)\approx 1 k_BT/R_H^3$ in the measurements, MCT fails to
describe $G''(\omega)$ adequately. If, however, the effective
packing fraction in the MCT calculations is adjusted to match the
loss modulus $G''(\omega)$, then this fit fails completely to
capture $G'(\omega)$ at high densities; see the dashed lines in
Fig.~\ref{Figadd}. Because the storage modulus, however, dominates
the linear mechanical response of the glass, the second fit needs to
be rejected. In conclusion, MCT correctly identifies the transition
to a glass at high densities with dominating elastic response and
yielding behavior under flow. It misses an additional dissipative
process, which contributes on the 10\% level to the shear moduli and
stresses in the frequency and shear rate window explored in our
experiments.

The existence of an additional dissipative process contradicts the
notion of 'ideal' glass states as described by the present ITT or
MCT aproach. Clearly, the system at $\phi_{\rm eff}=0.622$  becomes
a fluid at even longer times, or lower shear rates and frequencies
than observed in Fig.~\ref{Fig3}. This does not, however, contradict
the observation that the structural relaxation as captured in the
ITT equations has arrested. In an extension of the ITT approach, it
is possible to account for the additional decay channel in a an
extended schematic model; see the Appendix  for details. Results
from this extended F$_{12}^{(\gd)}$-model are included in
Fig.~\ref{Fig3} and demonstrate that none of the qualitative
features discussed within ITT change at finite frequencies or shear
rates. The additional process  leads to fluid behavior at even lower
$\omega$ or $\gd$, and needs to be taken into account only, if
exceedingly small frequencies or shear rates are tested; its
relaxation time at $\phi_{\rm eff}=0.622$ exceeds $10^8 R_H^2/D_0 =
8.8.10^3$ s. It does not shift the location of the 'glass
transition' as defined within the idealized ITT (MCT) approach,
because this is already determined by the shapes of the flow curves
and spectra in the observed windows.

\section{Conclusions}

In the present study, we explored the connection between the physics
of the glass transition and the rheology of dense colloidal
dispersions, including in strong shear flow. Using model colloidal
particles made of thermosensitive core-shell particles, we could
investigate in detail the vicinity of the transition between a
(shear-thinning) fluid and a (shear-molten) glass. The high
sensitivity of the particle radius to temperature enabled us to
closely vary the effective packing fraction around the critical
value. We combined measurements of the equilibrium stress
fluctuations, viz.~linear storage and loss moduli, with measurements
of flow curves, viz.~nonlinear steady state shear stress versus
shear rate, for identical external control parameters. In this way
we could verify the consequences from the recent suggestion, that
the glassy structural relaxation can be driven by shearing and in
turn itself dominates the low shear or low frequency rheology.

In the employed theoretical approach, the equilibrium structure as
captured in the equilibrium structure factor $S_q$ sufficed to
describe all phenomena qualitatively. As only exception, we observed
an ultra-slow decay of all glassy states that is yet not accounted
for by theory. Microscopic calculations were possible for the linear
response quantities using mode coupling theory applied to hard
spheres. Schematic model calculations were possible within the
integration through transients approach, and simultaneously captured
the linear and nonlinear rheology using identical parameter sets.
Semi-quantitative agreement between microscopic and schematic model
calculations and with the measured data for varying effective
packing fraction could be achieved adjusting a small number of fit
parameters in smooth variations.

\section{Acknowledgements}
We thank G. Petekidis and M. Cates for helpful discussions, and acknowledge
financial support by the Deutsche Forschungsgemeinschaft in IRTG 667
'Soft Condensed Matter Physics'. We acknowledge financial support by
the DFG, SFB 481, Bayreuth, and by the Forschergruppe 'Nonlinear
Dynamics of Complex Continua', Bayreuth.

\appendix

\section{Extended model including hopping}

The ITT equations contain the feed back mechanism that the friction
increases because of slow structural rearrangements. In the
schematic F$_{12}^{(\gd)}$-model this is captured by the
approximation for the generalized friction kernel $m(t)$ in \gl{d2}.
For $\gd=0$ it leads to non-ergodic glass states at large enough
vertices $v_{1,2}$. A dissipative process explaining the fluidity of
glassy states should renormalize the diffusion kernel $\Delta(t)$.
Moreover, this mechanism should become more important the longer the
relaxation time in $m(t)$. If, however, the additional dissipative
process is too strong, all effects of the bare ITT approach are
smeared out and the described phenomenology of the glass transition
can not be observed.

G\"otze and Sj\"ogren found when considering (possibly unrelated)
dissipative processes in simple liquids that this can by achieved by
splitting the diffusion kernel into two decay channels, one
connected to the original $m(t)$, and the other one connected to the
new dissipation mechanism. In order for the second decay channel to
take over in glassy states, it suffices to model it by one
additional parameter $\delta$ in a linear ansatz $\Delta^{\rm
dissip}(t) = \delta\, m(t)$. This leads to the following replacement
of \gl{d1} in the F$_{12}^{(\gd)}$-model
\begin{equation}\label{ap1}
\partial_t \Phi(t) + \Gamma \left\{\! \Phi(t) +\!\! \int_0^t\!\!\!\!dt'\, m(t\!-\!t') \, \left[
\partial_{t'} \Phi(t')\! + \delta \, \Phi(t') \right]\!  \right\}  = 0
\end{equation}
The memory function $m(t)$ is still given by \gl{d2} because
shearing decorrelates arbitrary fluctuations via shear-advection.
All parameters of the model are kept as specified in Sect.~IV, and
solutions of this extended model with parameter $\delta$ given in
the caption are included in Fig.~\ref{Fig3}. Importantly, the fluid
like behavior in the rheology at exceedingly small $\gd$ and
$\omega$ can now be captured without destroying the agreement with
the original ITT at higher parameters. Apparently, a single
parameter $\delta$ is not sufficient to model the  non-exponential
shape of the final relaxation process in the glass. Yet, further
extensions of the model in order to  describe this
non-exponentiality go beyond our present aim.

\section*{References}

\begin{thebibliography}{1}
\bibitem{eq5com}
Equation (\ref{b5}) presents a slightly simplified version of the
general expression in Ref.~\cite{Fuc:02}, restricting its use, so
that only variables $A$, like the shear stress $\sigma_{xy}$, with
$\langle A\rangle^{(\gd=0)}=0$ and $\langle A
\varrho_\qb\rangle^{(\gd=0)}=0$ can be considered in the following.
\bibitem{ishsm} Parameter $c^{(\gd)}$ is not yet well known as it could only
be estimated in the isotropically sheared hard sphere model (ISHSM),
which underestimates the effect of shear.\cite{Fuc:03,Bra:07}
\bibitem{erweiterung}
The schematic F$_{12}^{(\gd)}$-model of Refs.~\cite{Fuc:05a,Cra:06}
differs by setting $\gamma_c=1$.   This choice can be made by
rescaling $\Gamma$ and $v_\sigma$ if only a single experiment is
considered.\cite{Haj:07}
\bibitem{numerics} For the numerical solution of the schematic model
(\glto{d1}{d6}) the algorithm of Ref.~\cite{Fuc:91} was used, as has
been done in previous studies. \cite{Fuc:05a,Cra:06}  The
microscopic model (\glto{b7}{b11}) was evaluated without shear for
hard spheres with structure factor $S_k$ taken from the
Percus-Yevick approximation;\cite{russel} the wavevector integrals
were discretized according to Ref.~\cite{Fra:97} with only
difference that $M=600$ wavevectors were chosen from $k_{\rm
min}=0.05/R_H$ up to $k_{\rm max}=59.95/R_H$ with separation $\Delta
k = 0.1/R_H $. Time was discretized with initial step-width
$dt=2\,10^{-7} R_H^2/D_s$, which was doubled each time after 400
steps. Note that the MCT shear modulus at short times depends
sensitively on the large cut-off $k_{\rm max}$;\cite{Nae:98} for
hard spheres, \mbox{$g(t,\gd=0)=(n^2k_BT/60\pi^2) \int_{k_{\rm
min}}^{k_{\rm max}} dk\, k^4 (c'_k)^2 \; S_k^2\, \Phi_k^2(t)$} gives
the correct short time or high frequency divergence \mbox{$G'(\omega
\gg D_0/R_H^2) \sim \sqrt{\omega}$} only for $k_{\rm max}\to\infty$.
\end{thebibliography}



\end{document}